\documentclass[a4paper,latin9]{aa}
\usepackage{amsmath}
\usepackage{amssymb}
\usepackage{graphicx}
\usepackage{wasysym}
\usepackage{float}
\usepackage{placeins}

\usepackage{lipsum}
\usepackage{times}
\usepackage{multirow}
\usepackage{rotating}
\usepackage{longtable}
\usepackage{lscape}
\usepackage{epstopdf}
\defcitealias{pierre15}{Paper I}
\defcitealias{pacaud15}{Paper II}
\defcitealias{giles15}{Paper III}
\defcitealias{lieu15}{Paper IV}
\defcitealias{pompei15}{Paper VII}
\defcitealias{koulouridis15}{Paper XII}

\def\f#1   {Fig.~\ref{#1}}
\def\s#1   {Sect.~\ref{#1}}
\def\tab#1   {Tab.~\ref{#1}}
\def\eq#1   {Eq.~\ref{#1}}
\def\t#1   {Tab.~\ref{#1}}
\def\lum   {$L_{\mathrm{3~GHz}}$}
\def\comm#1   {{\tt (COMMENT: #1) }}

\def\Msol              {$\mathrm{M}_{\odot}$}
\def\msol              {$\mathrm{M}_{\odot}$~}

\def\msolyr              {$\mathrm{M}_{\odot}\, \mathrm{yr}^{-1}$ }

\def\wh                {W~Hz$^{-1}$}

\graphicspath{{Dropbox/Voronoi}}
\def\nVTAt {ten }
\def\nTott {seventeen }
\def\nXMM {7 }
\def\nXMMt {seven }
\def\nRadioInCluster {eight }

\titlerunning{A supercluster in the XXL-N field}
\authorrunning{Baran et al.}

\makeatother

\begin{document}
	
	\title{The XXL Survey: IX. Optical overdensity and radio continuum analysis of a supercluster at $z=0.43$\thanks{The full catalogue is available as a queryable database table XXL\_VLA\_15 via the XXL Master Catalogue browser http://cosmosdb.iasf-milano.inaf.it/XXL. A copy of the catalogue and the mosaic are is also available at the CDS via anonymous ftp to cdsarc.u-strasbg.fr (130.79.128.5).}} 
	
	\author{N. Baran\inst{1}, V. Smol{\v{c}}i{\'{c}}\inst{1},
		D. Milakovi{\'{c}}\inst{1}, M. Novak\inst{1}, J. Delhaize\inst{1}, F. Gastaldello\inst{2}, M.~E. Ramos-Ceja\inst{3}, F. Pacaud\inst{3}, \
		S. Bourke\inst{4}, C.~L. Carilli\inst{5}, S. Ettori\inst{6}, G. Hallinan\inst{4},
		C. Horellou\inst{7}, E. Koulouridis\inst{8}, L. Chiappetti\inst{2}, O. Miettinen\inst{1}, O. Melnyk\inst{1,11}, K. Mooley\inst{4}, M. Pierre\inst{8},
		E. Pompei\inst{9}, E. Schinnerer\inst{10}}
	
	\institute{University of Zagreb, Physics Department, Bijeni\v{c}ka cesta 32, 10000 Zagreb, Croatia
		\and INAF-IASF Milano, Via Bassini 15, I-20133 Milano, Italy
		\and Argelander-Institut f\"ur Astronomie, Auf dem H\"ugel 71, D-53121 Bonn, Germany
		\and California Institute Of Technology, Department of Astronomy, 1200 East California Boulevard, Pasadena, CA 91125, USA
		\and National Radio Astronomy Observatory, PO Box O, Socorro, New Mexico, 87801 USA
		\and INAF - Osservatorio Astronomico di Bologna, Via Ranzani 1, I-40127, Bologna, Italy
		\and Chalmers University of Technology, Department of Earth and Space Sciences, Onsala Space Observatory, 439 92 Onsala, Sweden
		\and Service d'Astrophysique AIM, DSM/IRFU/SAp, CEA-Saclay, F-91191 Gif sur Yvette, France
		\and European Southern Observatory, Alonso de C\'{o}rdova 3107, Vitacura, 19001 Casilla, Santiago 19, Chile
		\and Max Planck Institute for Astronomy, K\"onigstuhl 17, D-69117 Heidelberg, Germany
		\and Astronomical Observatory, Taras Shevchenko National University of Kyiv, 3 Observatorna St., 04053 Kyiv, Ukraine
	}
	
	\abstract{We present observations with the Karl G. Jansky Very Large Array (VLA)
		at 3~GHz (10~cm) toward a sub-field of the XXL-North 25~deg$^{2}$
		field targeting the first supercluster discovered in the XXL Survey.
		The structure has been found at a spectroscopic redshift of 0.43 and extending
		over $0^\circ\llap{.}35 \times 0^\circ\llap{.}1$ 
		on the sky. The aim of this paper
		is twofold. First, we present the 3~GHz VLA radio continuum observations, the
		final radio mosaic and radio source catalogue, and, second, we perform
		a detailed analysis of the supercluster in the optical and radio
		regimes using photometric redshifts from the CFHTLS survey and our
		new VLA-XXL data. Our final 3~GHz radio mosaic has
		a resolution of $3\farcs2\times1\farcs9$, and encompasses an
		area of $41\arcmin\times41\arcmin$ with rms noise level lower than $\sim20~\mu$Jy~beam$^{-1}$. The noise in the central $15\arcmin\times15\arcmin$ region is $\approx11\mu$Jy~beam$^{-1}$. From the mosaic we extract
		a catalogue of 155 radio sources with signal-to-noise ratio (S/N)$\geq6$, eight of which are large,
		multicomponent sources, and 123 ($79\%$) of which can be associated
		with optical sources in the CFHTLS W1 catalogue. Applying Voronoi
		tessellation analysis (VTA) in the area around the X-ray identified supercluster
		using photometric redshifts from the CFHTLS survey we identify a total
		of \nTott overdensities at $z_{\mathrm{phot}}=0.35-0.50$, \nXMM of
		which are associated with clusters detected in the \textit{XMM-Newton} XXL
		data. We find a mean photometric redshift of 0.43 for our overdensities,
		consistent with the spectroscopic redshifts of the brightest cluster
		galaxies of \nXMMt X-ray detected clusters. The full VTA-identified
		structure extends over $\sim0^\circ\llap{.}6 \times 0^\circ\llap{.}2$ on the
		sky, which corresponds to a physical size of $\sim12\times4$~Mpc$^{2}$ at $z=0.43$.
		No large radio galaxies are present within
		the overdensities, and we associate \nRadioInCluster (S/N$>7$) radio sources
		with potential group/cluster member galaxies. The spatial
		distribution of the red and blue VTA-identified potential group member
		galaxies, selected by their observed $g-r$ colours, suggests
		that the clusters are not virialised yet, but are dynamically young, as expected for hierarchical structure growth in a $\Lambda$CDM universe.
		Further spectroscopic data are required to analyse the dynamical state of the groups.}
	
	\maketitle
	
	\section{Introduction\label{sec:intro}}
	
	Over the last decade, deeper insights into various physical properties
	of galaxies, their formation and their evolution through cosmic time
	have been gained by sensitive, multiwavelength surveys such as GOODS
	\citep{dickinson03}, COSMOS \citep{scoville07}, GAMA (\citealt{driver09,driver11}) and CANDELS (\citealt{koekemoer11}; \citealt{grogin11}). The XXL\footnote{http://irfu.cea.fr/xxl
	} is a panchromatic survey (X-ray to radio) of two regions on the sky,
	each 25~deg$^{2}$ in size. Constraining the time evolution of the
	Dark Energy equation of state using galaxy clusters is the main goal
	of the XXL project. In a deep (6~Ms) \textit{XMM-Newton}%
	\footnote{http://xmm.esac.esa.int/ %
	} survey to a depth of $\sim5\times10^{-15}\, {\rm erg\, s^{-1}\, cm^{-2}}$
	in the {[}0.5-2{]}~keV band several hundred galaxy cluster detections
	($0<z<1.5$) were made over the $\sim$50~deg$^{2}$ \citep{pierre15} (hereafter \citet{pierre15}). The XXL South region (XXL-S) is located in the southern hemisphere
	in the Blanco Cosmology Survey (BCS) field%
	\footnote{http://www.usm.uni-muenchen.de/BCS/%
	} and the XXL North region (XXL-N) near the celestial equator encompasses
	the smaller XMM Large Scale Structure Survey (XMM-LSS)%
	\footnote{http://wela.astro.ulg.ac.be/themes/spatial/xmm/LSS/%
	} field. Multiwavelength photometric data ($<25$ in AB mag) from the
	UV/optical to the IR drawn from \textit{GALEX}, Canada-France-Hawaii Telescope
	Legacy Survey (CFHTLS), BCS, Sloan Digital Sky Survey (SDSS)%
	\footnote{http://www.sdss.org/%
	}, the Two Micron All Sky Survey (2MASS)%
	\footnote{http://www.ipac.caltech.edu/2mass/%
	}, \textit{Spitzer} Space Telescope%
	\footnote{http://ssc.spitzer.caltech.edu/%
	}, and Wide-field Infrared Survey Explorer (WISE)%
	\footnote{http://wise.ssl.berkeley.edu/%
	} already exist over almost the full area (\citet{pierre15}). Even deeper observations are available
	over smaller areas (\citet{pierre15}), and will be conducted over the full area with the DECam\footnote{http://www.darkenergysurvey.org/DECam/camera.shtml} to magnitude $25-26$ in \textit{griz} bands, and $26-27$ in $grizY$ bands with Hyper-SuprimeCam\footnote{http://www.naoj.org/Projects/HSC/}.
	Photometric redshifts are expected to reach accuracies better
	than $\sim10$\%, sufficient for large-scale structure studies out to intermediate redshift (z$\sim0.5$), and for evolutionary
	studies of galaxies and AGN to $z\sim3$. More than 15,000 optical spectra
	are already available (see \citet{pierre15}).
	
	To complement the multiwavelength coverage of the field, we present new
	radio observations with the Karl G. Jansky Very Large Array (VLA)
	at 3~GHz over a $\sim0^\circ\llap{.}7 \times 0^\circ\llap{.}7$ subarea of the
	XXL-N field. These observations target the first supercluster discovered in the
	XXL survey by Pompei et al. (hereafter \citet{pompei15}). One cluster within this large structure was previously identified at $z_{\mathrm{phot}}=0.48$ using photometric
	redshifts in the CFHTLS wide field by \citet{durret11}. 
	The \textit{XMM-Newton} XXL observations revealed
	six X-ray clusters within $\sim0^\circ\llap{.}35 \times 0^\circ\llap{.}1$
	as described in \citet{pierre15}, see also \citet{pacaud15} (hereafter \citet{pacaud15}). Based on further spectroscopic observations of the brightest cluster galaxies by \citet{koulouridis15} (hereafter \citet{koulouridis15}), \citet{pompei15} reports the redshift of the structure of $z=0.43$. The structure, thus, has a physical extent of $\sim10\times~2.9$~Mpc$^{2}$ ($21\arcmin\times6\arcmin$) . Here we present a detailed analysis
	of this structure based on the CFHTLS optical data and our new VLA
	radio data.
	
	Throughout this paper we use cosmological parameters in accordance with the \textit{Wilkinson} Microwave Anisotropy Probe satellite final data release (WMAP9) combined with a set of baryon acoustic oscillation measurements and constraints on $H_{0}$ from Cepheids and type Ia supernovae \citep{hinshaw13}. These parameters are $\Omega_{M}=0.282$, $\Omega_{\Lambda}=0.718$ and $H_{0}=69.7~\rm km~s^{-1}~Mpc^{-1}$. All sizes given in this paper are physical. All magnitudes are in the AB magnitude system, and all coordinates in J2000 epoch.
	
	In Sect.~\ref{sec:data} we present the optical and radio data used
	in this paper. In Sect.~\ref{sec:supercluster} we describe optical
	and radio properties of the supercluster. Discussion of the results
	is presented in Sect.~\ref{sec:discussion} and the summary is given
	in Sect.~\ref{sec:summary}. 
	
	\section{Data}
	
	\label{sec:data}
	
	\subsection{Optical data}
	
	\label{sub:optical}
	
	We use the photometric redshift catalogues from the CFHTLS%
	\footnote{http://www.cfht.hawaii.edu/Science/CFHTLS/%
	}. The CFHTLS Wide survey is divided into fields W1, W2, W3 and W4
	covered by the $u${*}, $g$', $r$', $i$', $z$', and $y$' bands with an 80\% completeness limit
	in AB magnitudes for point sources of $u${*}=25.2, $g$'=25.5, $r$'=25.0,
	$i$'=24.8 and $z$'=23.9. The XXL-N field lies within the CFHTLS W1 field.
	The photometric redshift accuracy in this field is $\sigma_{\Delta{z}/(1+z_{s})}\sim0.030-0.032$
	at $i<21.5$, $\sigma_{\Delta{z}/(1+z_{s})}\sim0.066-0.077$
	at $22.5<i<23.5$, where $\Delta{z}=|z_{p}-z_{s}|$ is the difference between
	photometric and spectroscopic redshift as described in \citet{ilbert06} and \citet{coupon09} (see also catalogue documentation)\footnote{\tiny{http://cesam.lam.fr/cfhtls-zphots/files/ cfhtls\_wide\_T007\_v1.2\_Oct2012.pdf}}.
	Since the redshift accuracy rapidly deteriorates for magnitudes above $i=23.5$ in the optical analysis of the structure, we have adopted a cut in $i$-band magnitude at $i=23.5$.
	
	\begin{figure*} 
		\centering 
		\includegraphics[scale = 0.56]{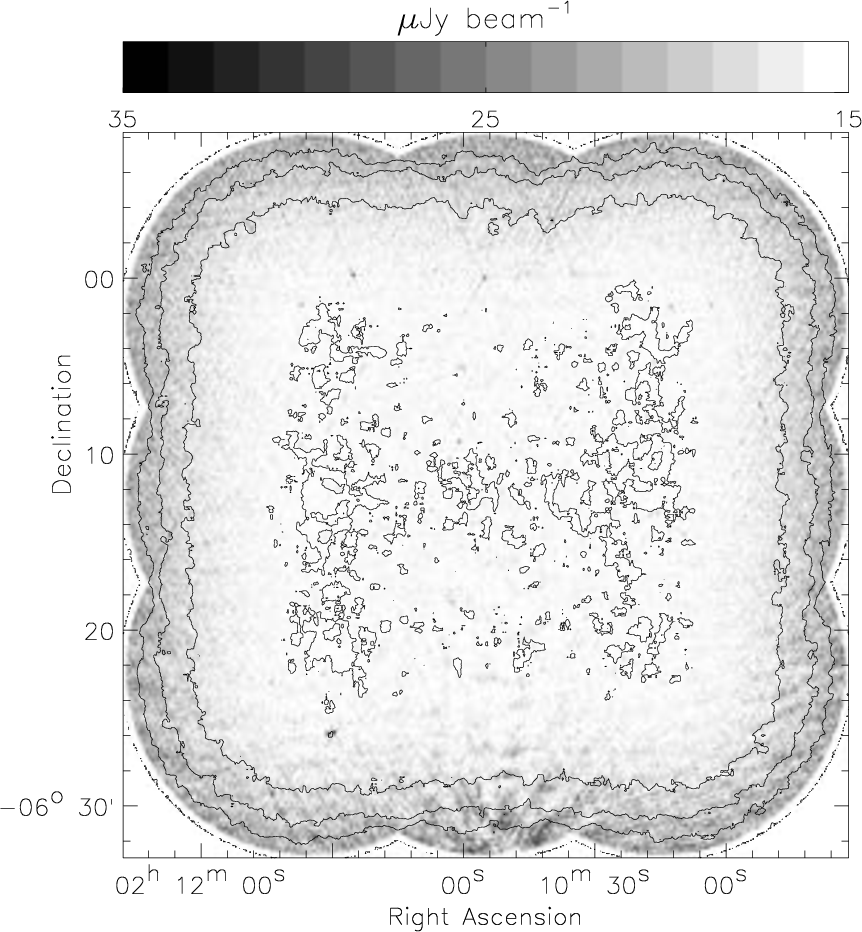}\\
		\includegraphics[scale = 0.5]{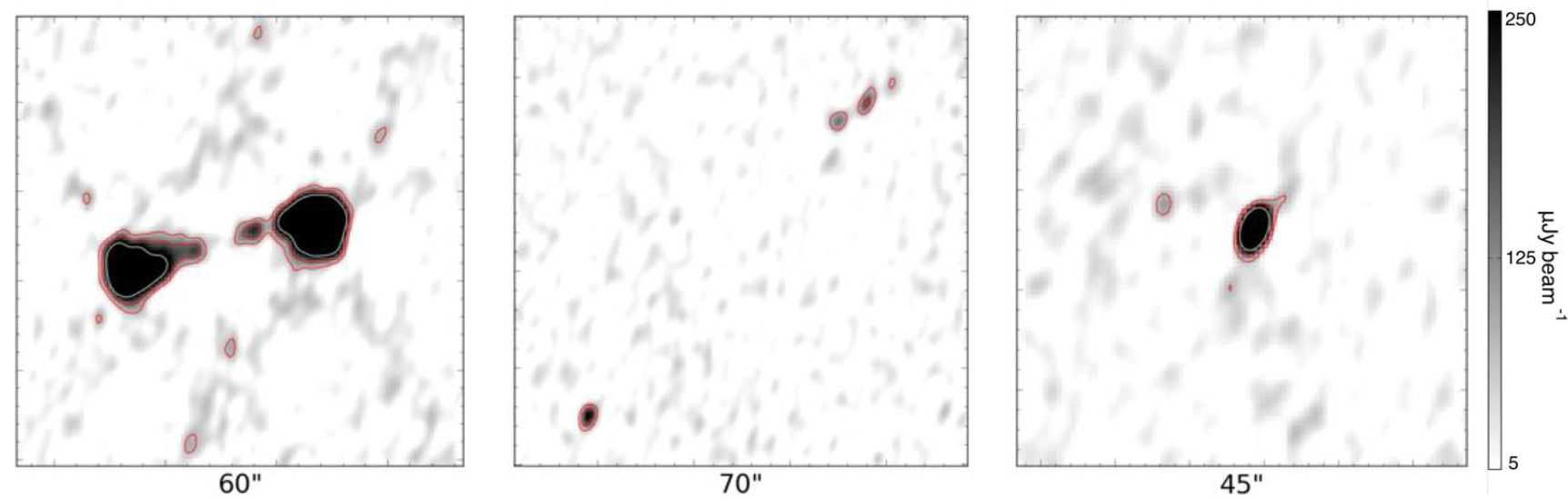}
		\protect\caption{\textit{Top}: Greyscale mosaic of the area of XXL-N field observed with the VLA at 3~GHz with rms contours overlaid (contour levels are 10, 20, 30 and 40 $\mu$Jy~beam$^{-1}$). \textit{Bottom:} Three examples of single- and multicomponent sources represented as greyscale maps with contours overlaid. The contour levels are 2$^i$~$\sigma$ ($i$~=~2,3, and 4, 1$\sigma$~=~11$~\mu$Jy~beam$^{-1}$).} 
		\label{fig:mfmap} 
	\end{figure*}
	
	To obtain stellar mass estimates used in Section~\ref{sec:discussion} we used publicly available data from the Canada-France-Hawaii Lensing Survey (CFHTLenS) (see \citealt{heymans12}, \citealt{hildebrandt12}, \citealt{erben13}, and \citealt{velander12}). The stellar masses were extracted using ''LePhare'' (see \citealt{ilbert06} and \citealt{arnouts99}) assuming a Chabrier initial mass function (see \citealt{chabrier03}).
	
	\subsection{VLA 3~GHz radio data}
	\label{sec:radio} 
	\subsubsection{Observations, data reduction, and imaging}
	\label{sec:reduction} We observed nine regularly spaced pointings
	separated by 10$'$ in RA and Dec in the XXL-N field with the VLA
	in B configuration, as given in the Table~\ref{tab:pointings}. The area observed has central coordinates 
	RA$~=~\rm 02^h10^m54^s\llap{.}9, \ Dec~=~-06^\circ12^\prime20^{\prime\prime}\llap{.}2$.
	Three hours of observations were taken on August 30, 2012 (7:35 $-$
	10:35 UTC) in S-band with central frequency of 3~GHz and 2~GHz bandwidth
	divided into 16 spectral windows 128~MHz in width. Owing to large-amplitude
	radio frequency interference (RFI) between 2.12-2.38~GHz, the second
	and third spectral windows were completely flagged. For flux calibration
	we observed the calibrator 0521+166 (Seyfert 1 galaxy 3C138) for 105
	seconds at the end of the observations. For phase calibration we used
	J0241-0815 (Seyfert 2 galaxy NGC1052). The
	phase calibrator was observed every 30 minutes for 2 minutes on-source
	throughout the observations. The observations were taken under a clear
	sky at a temperature of 12$\,^{\circ}$C.
	
	\begin{table}
		\caption {
			List of VLA pointings used in the observations.}
		\begin{tabular}{ c c c c }
			\hline\hline
			Pointing & RA & Dec & Time on-source \\
			&	{[}h m s{]} 		   & {[}$^\circ~\arcmin~\arcsec${]}		 & [s] \\
			(1)&(2)&(3)&(4)\\
			\hline
			P1       & 02 10 14.66      & -06 22 20.00 & 972\\
			P2       & 02 10 14.66      & -06 12 20.00 & 987\\
			P3       & 02 10 14.66      & -06 02 20.00 & 975\\
			P4       & 02 10 54.89      & -06 22 20.00 & 984\\
			P5       & 02 10 54.89      & -06 12 20.00 & 981\\
			P6       & 02 10 54.89      & -06 02 20.00 & 975\\
			P7       & 02 11 35.11      & -06 22 20.00 & 999\\
			P8       & 02 11 35.11      & -06 12 20.00 & 984\\
			P9       & 02 11 35.11      & -06 02 20.00 & 978\\
			\hline
		\end{tabular}
		\tablefoot{Pointing names are stated in column (1), and their central coordinates in sexagesimal format in columns (2) and (3). Total on-source time in seconds is given in column (4). All observations were performed in the S-band with 3~GHz central frequency; B configuration has the largest recoverable angular scale of $58\arcsec$.}
		\label{tab:pointings}
	\end{table}
	
	Calibration of the data was done using the \textsc{AIPSLite} pipeline (e.g.
	\citealt{bourke14}) developed for the
	Caltech-NRAO Stripe 82 Survey \citep{mooley15}.
	The calibrated data were then imported into the \textsc{CASA} package and data with amplitudes greater than 0.4~Jy were further flagged using the \textsc{CASA}
	task FLAGMINMAX. Using the task SPLIT
	the $uv$ data were then split into separate pointings prior to imaging.
	Imaging of the $uv$ data was performed for
	each pointing separately using multifrequency synthesis within the
	\textsc{CASA} task CLEAN (\citealt{rau11}; see also \citealt{novak15}).
	The algorithm models the sky brightness in the entire 2~GHz bandwith
	using a linear combination of Gaussian functions with amplitudes
	following a Taylor polynomial in frequency. A two-term expansion
	was used with 10,000 iterations, threshold 3$\sigma$ (40~$\mu$Jy),
	default gain value of 0.1 and `gridmode' parameter set to wide field
	with 128 \textit{w}-projection planes for convolution. We used Briggs weighting
	with a robust value of 0.5. This value was chosen as it optimally balances
	rms and sidelobe contamination. After the imaging, a wide-band
	primary beam correction was applied using the \textsc{CASA} task WIDEBANDPBCOR.
	This was executed with two Taylor terms (`nterms') and
	a threshold of 20\% of the primary beam response was set.
	
	Each pointing was convolved to $3\farcs2\times1\farcs9$ with position angle of $20^\circ$ to assure a uniform synthesised beam across the final mosaic. To construct the mosaic we combined the nine individual pointings by performing an rms-weighted sum of the pixels in the overlapping regions. For each pixel the local rms value was obtained from the corresponding pixel in the noise map. These noise maps were constructed using the \textsc{AIPS}\footnote{http://www.aips.nrao.edu/} task RMSD and are discussed in more detail in Sect.~\ref{sec:catalogueing}. 
	
	In summary, the final $3\farcs2\times1\farcs9$ resolution mosaic shown in Fig.~\ref{fig:mfmap} covers a total area of $41\arcmin\times41\arcmin$ ($0.47$~deg$^{2}$) and has an average rms of $20~\mu$Jy~beam$^{-1}$. The average rms of the inner $15\arcmin\times15\arcmin$ region is $10.8~\mu$Jy~beam$^{-1}$. Within this central area the noise is nearly Gaussian.
	
	\subsubsection{Cataloguing}
	\label{sec:catalogueing} 
\begin{figure}[H]
		\centering \includegraphics[bb=0 0 600 604,width=1\columnwidth]{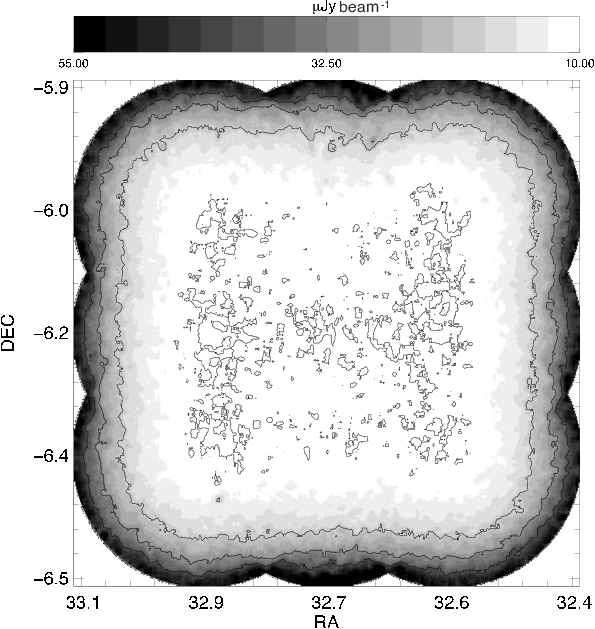}
		\protect\caption{Noise map of VLA - XXL-N mosaic generated using the \textsc{AIPS} task RMSD.
			The rms contour levels are at 10, 20, 30 and 40 $\mu$Jy~beam$^{-1}$.
			The grey scale (in units of $\mu$Jy~beam$^{-1}$) is shown at the top.}
		\label{fig:mfmaprms} 
	\end{figure}

To catalogue all sources present in the mosaic
	down to a specific signal-to-noise ratio (S/N) we used the \textsc{AIPS} task
	Search and Destroy (SAD). The SAD task looks for potential sources in an image
	by identifying agglomerations of pixels (islands) with values above
	a certain level and then fits Gaussian components to them (see \textsc{AIPS}
	CookBook%
	\footnote{http://www.aips.nrao.edu/cook.html%
	}). We ran SAD using both the mosaic
	and the rms maps as inputs. The latter
	was generated via the \textsc{AIPS} task RMSD and is used by SAD to determine
	the local S/N values (see Fig.~\ref{fig:mfmaprms}). The size of the box from which the rms
	is sampled (parameter `imsize') was set to 80$\times$80
	pixels and parameters `xinc' and `yinc',
	which define the increments for calculating rms, were set to 10
	pixels. The task SAD was configured with the following parameters:
	`ngauss' is the maximum number of components to look
	for was set to 10,000, `cparm' defines the S/N levels down
	to which the task will look for potential sources and was set to
	50 10 7 4.5, `dparm(9)' was set to 2.1 to set the
	units of `cparm' levels to S/N, `doall'
	was set to -1 to fit only one Gaussian per island. The parameter
	`dparm(7)', which defines total residual flux in the
	fitting box, was set to 85~$\mu$Jy. All other parameters were kept at default
	values. This procedure yielded 311 radio source components down to 4.5 times the
	local rms, all of which were visually inspected
	and compared to CFHTLS optical images of field W1 with a 1$\arcsec$
	position cross-match radius. 
	
	A total of 28 components were manually combined into eight multicomponent sources. Their total integrated flux densities were measured using the \textsc{AIPS} task TVSTAT down to 2~$\sigma$ contours
	(while their peak surface brightness values, major and minor axes, position angles, and their respective errors are set to -99.99 values in the catalogue;
	see below). In all such cases the coordinates of the component that
	overlapped with the optical detection were taken as the source position. 
	
	The fraction of radio and optical matches (within 1\arcsec matching radius) is roughly constant at around 80\% down to 6$\sigma$, and then drops to 56\% for $5<$S/N$<6$. Therefore, in the final catalogue we report only 155 sources above 6$\sigma$.
	
\begin{figure}[H]
		\centering \includegraphics[scale = 0.5]{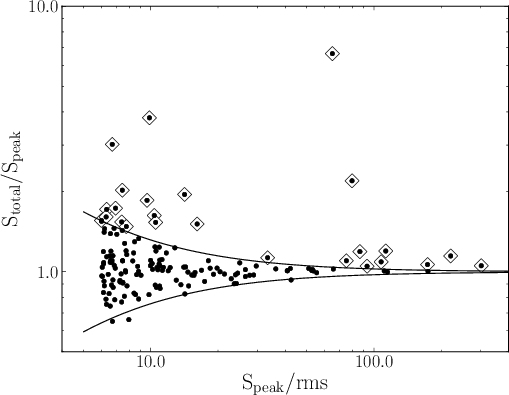} \protect\caption{Ratio of 3~GHz total flux densities and peak surface brightnesses as a function of the
			signal-to-noise ratio. Sources above (below) the upper envelope are considered resolved (unresolved). Resolved sources are indicated by diamonds. See Sect.~\ref{sec:catalogueing} for details.}
		\label{fig:resolvedunresolved} 
	\end{figure}

	To separate resolved from unresolved sources we made use of
	the ratio of total flux density and peak surface brightness, which is a measure of extendedness
	of the sources. The ratio $S_{\textrm{total}}/S_{\textrm{peak}}$
	as a function of the S/N is shown in Fig.~\ref{fig:resolvedunresolved},
	and spreads with decreasing S/N (see, e.g. \citealt{bondi08}). Following the method of \citet{bondi08}, we fitted a lower envelope containing $\simeq$~95\% of the sources with $S_{\textrm{peak}}>S_{\textrm{total}}$. This envelope was then mirrored above the line defined by $S_{\textrm{total}}/S_{\textrm{peak}}=1$, giving an equation for the upper envelope
	\begin{align}
	\frac{S_{\textrm{total}}}{S_{\textrm{peak}}}=1+\frac{4}{(\frac{S_{\textrm{peak}}}{rms})^{1.1}}.
	\end{align}
	
	Sources in Fig.~\ref{fig:resolvedunresolved} above the upper envelope are considered resolved. For unresolved sources the values of total flux densities are then set equal to their peak surface brightnesses. Lengths of major and minor axes, and position angles (with their respective errors) are set to 0.0.
	
	A total of 25 (130) sources were considered
	resolved (unresolved). A sample page of the catalogue is given in Table
	\ref{tab:cattable}. 
	The Table~\ref{tab:cattable} shows a sample radio catalogue page adn the relevant parameters.
	In summary, 155 sources are catalogued, 25 of which are considered resolved, 8 are flagged as multicomponent, and 11 are positioned on a sidelobe.
	
	\subsubsection{Multiwavelength properties of the 3~GHz sources}
	
	We found 123 optical counterparts of the 155 3~GHz sources using the
	CFHTLS W1 field catalogue and a 1$\arcsec$ matching radius, corresponding to 79\% of the sources. All sources
	with optical counterparts are visible across all CFHTLS bands ($u${*},$g$,$r$,$i$,$z$).
	For comparison, a correlation of 155 randomly distributed artificial
	sources across the field yielded nine matches with the CFHTLS catalogue
	within 1$\arcsec$ matching radius. This sets the false-match probability
	to $\lesssim6\%$.
	
	The $i$-band magnitudes of the matched sources are in the range $\sim15$ to $\sim25$ with an average $\bar{i}\approx21.3$.
	Redshifts range from $\sim0.3$ to $\sim1.2$ and have an average of $\bar{z}_{phot}\simeq0.8$.
	
	We obtained 1.4~GHz information for our 3~GHz sources using the
	Faint Images of the Radio Sky at Twenty-cm (FIRST
	\footnote{http://sundog.stsci.edu/
	}). The FIRST survey is an all-sky survey conducted at 1.4~GHz; it has
	angular resolution of $5\arcsec$ \citep{becker95} and a typical rms of 0.15~mJy. 
	For comparison, the average beam size in the VLA-XXL survey was $\sim~2\farcs6$ with an average rms of $\sim~20~\mu$Jy, giving $\sim4$ times better sensitivity (for an assumed spectral index of $\alpha~=~-0.8$) and a factor of $\sim2$ better resolution. The FIRST source catalogue lists sources brighter than $0.75$~mJy. 
	
	By positionally matching our 3~GHz source catalogue with the FIRST Survey Catalog \citep{helfand15} using a matching
	radius of $2\farcs5$ we find 21 matches. Of these 21 sources, 16 are singlecomponent at 3~GHz. We find a mean spectral
	index of $\alpha=-0.57$ with a standard deviation of $\sigma=0.51$ between the 3 and 1.4~GHz fluxes for these singlecomponent sources (assuming $S_{\nu}\propto\nu^{\alpha}$, where $S_{\nu}$ is the radio
	flux density at frequency $\nu$, and $\alpha$ is the spectral index). These values are expected for radio sources
	at these flux levels (e.g. \citealt{kimball08}).
	
	\begin{table*}
		\centering \protect\caption{Sample radio catalogue page.}
		\tabcolsep=0.11cm
		\begin{tabular}{c c c c c c c c c c c c c c c c c c}
			\hline\hline
			ID & RA & Dec & $\sigma_{\textrm{{\scriptsize RA}}}$ & $\sigma_{\textrm{{\scriptsize Dec}}}$ & S$_{\textrm{{\scriptsize peak}}}$ & $\sigma_{\textrm{{\scriptsize peak}}}$ & S$_{\textrm{{\scriptsize total}}}$ & $\sigma_{\textrm{{\scriptsize total}}}$ & \scriptsize Res & \scriptsize SL & \scriptsize Multi & Maj & $\sigma_{\textrm{{\scriptsize Maj}}}$ & Min & $\sigma_{\textrm{{\scriptsize Min}}}$ & PA & $\sigma_{\textrm{{\scriptsize PA}}}$ \\
			& {\tiny{[}h m s{]}} & {\tiny{[}$^\circ~\arcmin~\arcsec${]}} & {\tiny{[}\arcsec{]}} & {\tiny{[}\arcsec{]}} & {\tiny{[}$\mu$Jy~beam$^{-1}${]}} & {\tiny{[}$\mu$Jy~beam$^{-1}${]}} & {\tiny{[}$\mu$Jy{]}} & {\tiny{[}$\mu$Jy{]}} &  &  &  & {\tiny{[}\arcsec{]}} & {\tiny{[}\arcsec{]}} & {\tiny{[}\arcsec{]}} & {\tiny{[}\arcsec{]}} & {\tiny{[}$^\circ${]}} & {\tiny{[}$^\circ${]}} \\
			(1)&(2)&(3)&(4)&(5)&(6)&(7)&(8)&(9)&(10)&(11)&(12)&(13)&(14)&(15)&(16)&(17)&(18)\\ 
			\hline
			1  & 02 09 41.78 & -06 19 10.56 & 0.18 & 0.43 & 199  & 30 & 601  & 42  & 1 & 0 & 0 & 6.7 & 1.0 & 2.7 & 0.4 & 172.4 & 6.0\\
			2  & 02 09 42.38 & -05 59 42.66 & 0.05 & 0.08 & 494  & 29 & 494  & 29   & 0 & 0 & 0 & 0.0 & 0.0 & 0.0 & 0.0 & 0.0   & 0.0\\
			3  & 02 09 45.62 & -06 18 43.14 & 0.02 & 0.03 & 1229 & 23 & 1229 & 23   & 0 & 0 & 0 & 0.0 & 0.0 & 0.0 & 0.0 & 0.0   & 0.0\\
			4  & 02 09 47.75 & -06 20 53.02 & 0.15 & 0.19 & 134  & 20 & 134  & 20   & 0 & 0 & 0 & 0.0 & 0.0 & 0.0 & 0.0 & 0.0   & 0.0\\
			5  & 02 09 48.68 & -06 29 41.37 & 0.09 & 0.12 & 453  & 42 & 453  & 42   & 0 & 0 & 0 & 0.0 & 0.0 & 0.0 & 0.0 & 0.0   & 0.0\\
			6  & 02 09 51.03 & -06 06 20.77 & 0.02 & 0.03 & 928  & 18 & 928  & 18   & 0 & 0 & 0 & 0.0 & 0.0 & 0.0 & 0.0 & 0.0   & 0.0\\
			7  & 02 09 51.90 & -06 07 12.60 & 0.12 & 0.26 & 158  & 16 & 602  & 52   & 1 & 0 & 0 & 6.1 & 0.6 & 2.6 & 0.3 & 10.3  & 4.0\\
			8  & 02 09 54.42 & -05 54 54.17 & 0.13 & 0.22 & 184  & 30 & 184  & 30   & 0 & 0 & 0 & 0.0 & 0.0 & 0.0 & 0.0 & 0.0   & 0.0\\
			9  & 02 09 56.19 & -06 09 16.01 & 0.07 & 0.09 & 235  & 15 & 356  & 56   & 1 & 0 & 0 & 3.4 & 0.2 & 2.7 & 0.2 & 163.7 & 11.0\\
			10 & 02 09 56.43 & -06 04 57.98 & 0.20 & 0.27 & 80   & 13 & 125   & 32   & 1 & 0 & 0 & 4.1 & 0.7 & 2.3 & 0.4 & 151.7 & 11.0\\
			\hline\end{tabular}
		\tablefoot{ Column (1) gives the ID of the source from
			the VLA radio catalogue, columns (2) and (3) represent the coordinates of the source. Position errors in arcseconds, as calculated using the \textsc{AIPS} task SAD, are shown in columns (4) and (5). Peak surface brightness and its error are given in columns (6) and (7) in units of $\mu$Jy~beam$^{-1}$, followed by total flux density in column (8) and its error in column (9), both given in $\mu$Jy. A flag denoting a resolved source is stated in column (Res = 1 it the source is resolved, 0 otherwise)(10), while a flag for a source located on a sidelobe is given in column (11). Sources found on a sidelobe by visual inspection were flagged with Sidelobe = 1. A flag indicating a source composed of multiple components is stated in column (12) (Multi = 1 if the source is multicomponent, 0 if single component). The major axis FWHM of the deconvolved Gaussian component and its error are given in columns (13) and (14) in arcseconds, respectively. Column (15) gives the minor axis FWHM of the fitted Gaussian with its error in column (16) in arcseconds. The position angle of the major axis of the fit Gaussian and its error are shown in (17) and (18) in degrees, respectively. In total, 25 sources are considered resolved, eight have been flagged as multicomponent and 11 as positioned on a sidelobe. The full catalogue is available as a queryable database table XXL\_VLA\_15 via the XXL Master Catalogue browser \textit{http://cosmosdb.iasf-milano.inaf.it/XXL}. A copy is also available at the CDS via anonymous ftp to cdsarc.u-strasbg.fr (130.79.128.5).}
		\label{tab:cattable} 
	\end{table*}
	
	\section{Analysis of the supercluster}
	
	\label{sec:supercluster}
	
	In this section we analyse the first supercluster (XLSSC-e) identified in the
	XXL-N field (\citet{pompei15}). We use Voronoi tesselation analysis (VTA) and the
	CFHTLS photometric redshifts (Sect.~\ref{sub:voronoi}), as well as
	the radio properties of the VTA-identified potential supercluster
	members (Sect.~\ref{sub:radioproperties}).
	
	\begin{figure}[H]
		\includegraphics[scale = 0.5]{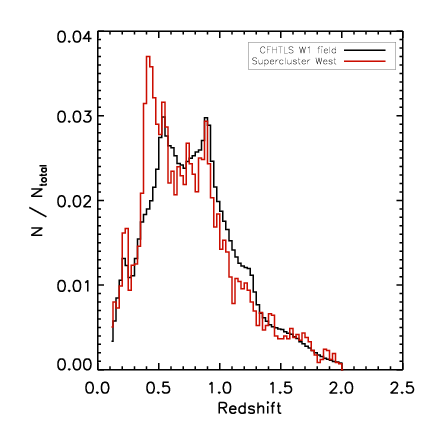} \protect\caption{Normalised redshift distribution of galaxies in CFHTLS field W1 (black)
			and the region of the XXL-N field containing the six westernmost X-ray
			identified clusters (red). }
		\label{fig:zdistr} 
	\end{figure}
	
	\subsection{Optical properties}
	
	\label{sub:voronoi}
	
	\subsubsection{Identification of overdensities:  Voronoi tessellation analysis}
	\label{subsub:vta}
\begin{figure*}[!]
		\includegraphics[width = 1\linewidth]{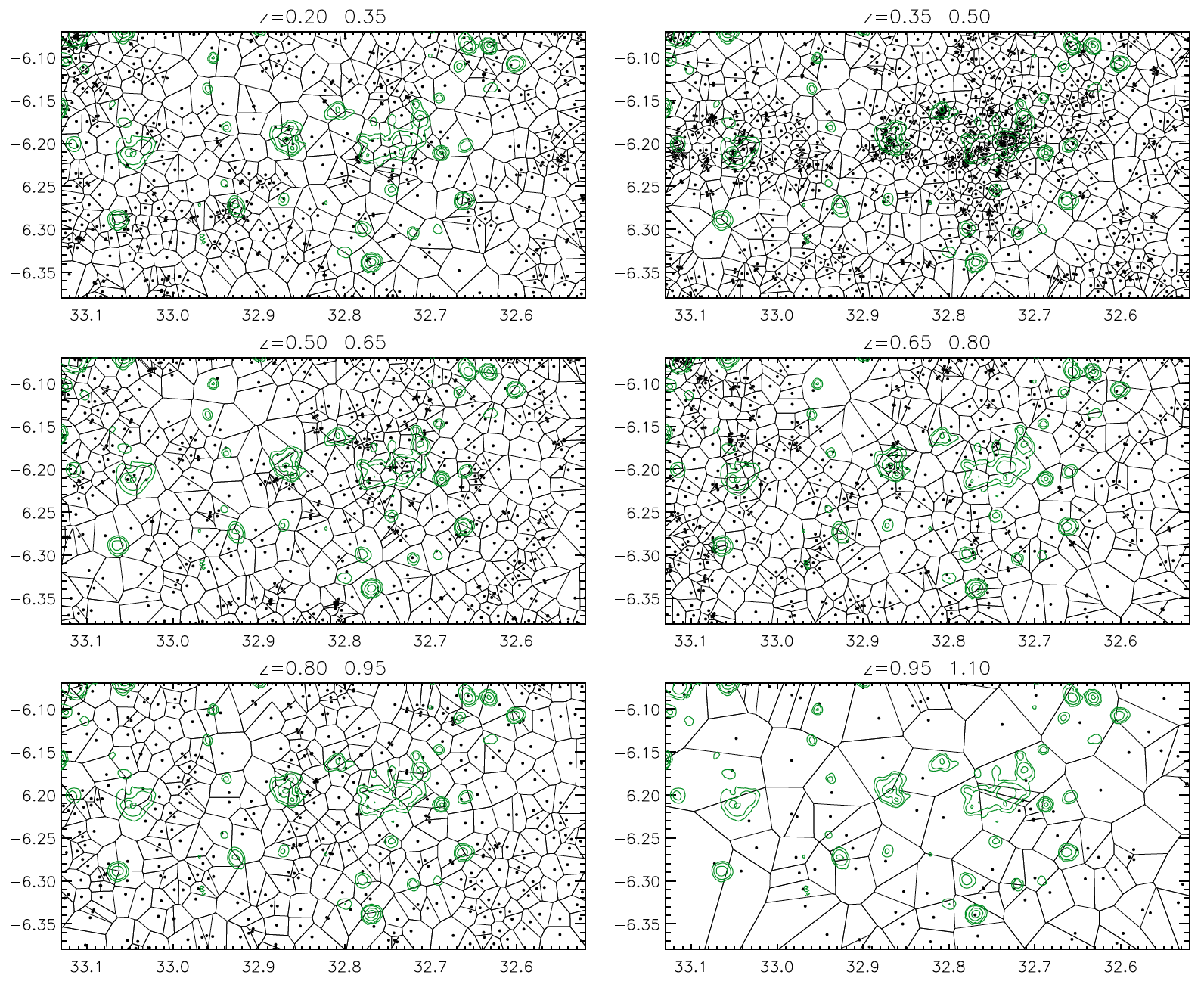}
		\protect\caption{Voronoi tessellation diagrams of the region encompassing the five
			westernmost X-ray clusters (RA 32.52, 33.13, Dec: -6.38,-6.05) as a function of redshift ($0.2<z<1.1$)
			with Voronoi polygons shown with lines and isophotes of X-ray surface brightness
			in contours. Each polygon contains only one galaxy, indicated by a black
			dot. The largest number of sources in the region of X-ray emissions are located in the redshift range $0.35<z<0.50$. The levels of the X-ray contours are shown in green at 5.28, 7.02, 12.54 and 30 cts~s$^{-1}$~deg~$^{-2}$ in the [0.5-2.5]~keV range.}
		\label{fig:z_voronoi} 
	\end{figure*}

	To identify the photometric redshift depth that the supercluster
	occupies, in Fig.~\ref{fig:zdistr} we compare the photometric redshift distribution data from
	the CFHTLS T0007 Data Release (described in \citealt{ilbert06, coupon09}, see also catalogue documentation\footnote{http://cesam.lam.fr/cfhtls-zphots/files/cfhtls\_wide\_T007\_v1.2\_Oct2012.pdf}), between the area encompassing the six westernmost X-ray clusters and the entire field. The comparison indicates an excess in the
	number of galaxies at $z\approx0.43$ in this area. In Fig.~\ref{fig:z_voronoi}
	we show the spatial distribution of galaxies in this cluster area for six
	$\Delta z=0.15$ wide redshift bins in the range $z_{\mathrm{phot}}=0.2-1.1$.
	Consistent with the redshift peak at $z\sim0.43$ we find the strongest
	clustering of galaxies in the redshift bin $z_{\mathrm{phot}}=0.35-0.50$ (cf. \citet{coupon09}). This range corresponds to a width of $3.4\sigma_\mathrm{\Delta z/(1+z)} (1+z)$ at z=0.43, and we consider this redshift range in the analysis. 
	
\begin{figure}[!]
		\includegraphics[scale = 0.5]{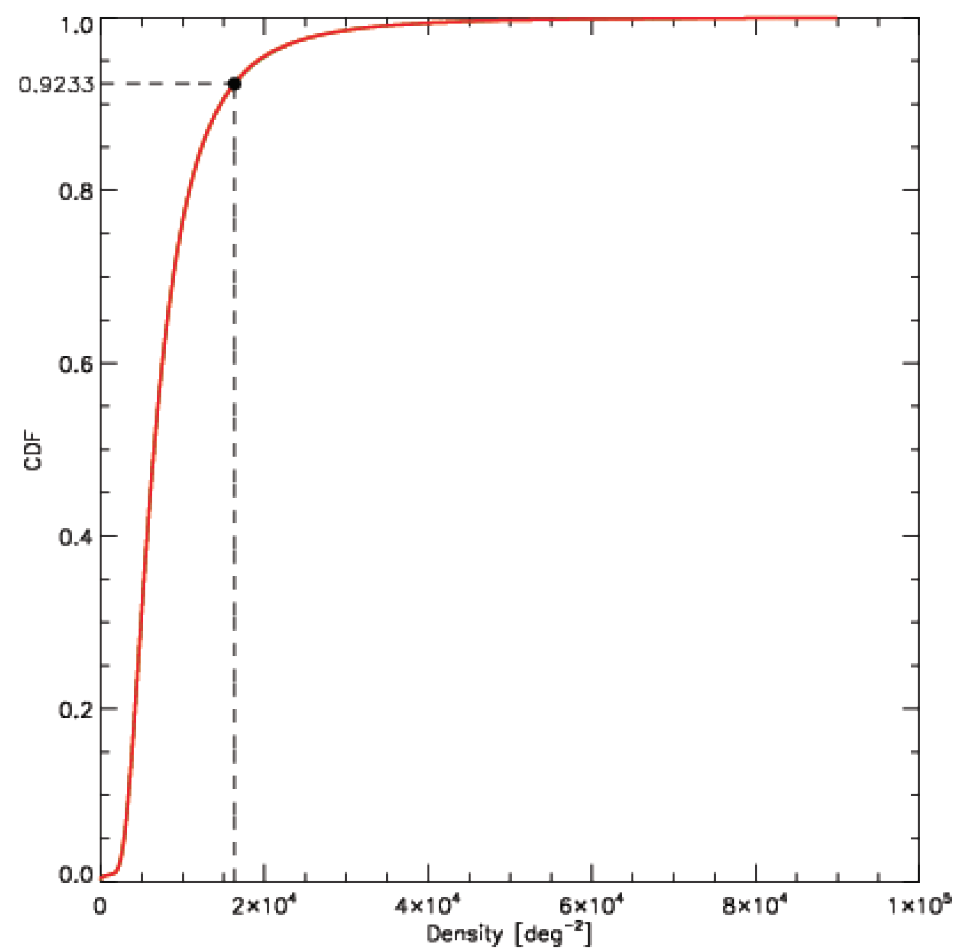}
		\protect\caption{Cumulative distribution function (CDF) of galaxy densities in ten simulated catalogues with randomly distributed artificially generated galaxies. The overdensity threshold used in this paper corresponds to the 92\% quantile of the distribution. Galaxies populating Voronoi cells with density above this value are treated as overdense region/cluster member candidates.}
		\label{fig:cdf} 
	\end{figure}

	To identify potential supercluster galaxy members we perform a VTA in the area of interest for CFHTLS W1
	galaxies with $z_{\mathrm{phot}}=0.35-0.50$. In this technique (used in \citealt{icke89}, and also recently in e.g. \citealt{smolcic07}; \citealt{oklopcic11}; \citealt{jelic12}) we divide the area into polygons and calculate their
	surface area. Each polygon contains only one galaxy with sides equidistant from the nearest neighbouring galaxies and
	from the galaxy within. This means the inverse of
	the surface of these polygons, $A^{-1}$, is proportional to the local
	galaxy density $\rho_{\textrm {local}}$, i.e. $\rho_{\textrm {local}}\propto A^{-1}$.
	A robust `local' galaxy density threshold value then identifies
	`overdense' regions. To determine this threshold we performed the
	VTA on an area $\sim10$ times larger than the area of interest. This area was populated by $N$ randomly distributed artificially generated galaxies. The number of galaxies $N$ is equal to the number of galaxies with $z_{\mathrm{phot}}=0.35-0.50$ within the CFHTLS catalogue in the same area.
	Based on the VTA on this field we calculated the average field number
	density, i.e. the median of the $A_{i}^{-1}$ distribution where $A_{i}$
	is the area of a given Voronoi cell. We note that the CFHTLS images of this region contained masked areas, for example around bright stars or artefacts. The impact of these masks on our density estimates was compensated for by scaling the overdensity criterion threshold by a factor equal to $(A_{\textrm {Total}}-A_{\textrm {Masked}})/A_{\textrm {Total}}$, which equates to $\sim$93\%. The threshold was chosen as the 92\% quantile of the cumulative distribution of galaxy densities in the mock catalogues (see Fig.~\ref{fig:cdf}) and equates to 16,357 galaxies per square degree. We then flagged as overdense those cells for which $A_{i}^{-1}$ is greater than this threshold. Galaxies within such flagged cells are considered to be potential members of overdense structures. 
	
\begin{figure*}[!]
		\centering \vspace{1cm}
		\includegraphics[width = 1\linewidth]{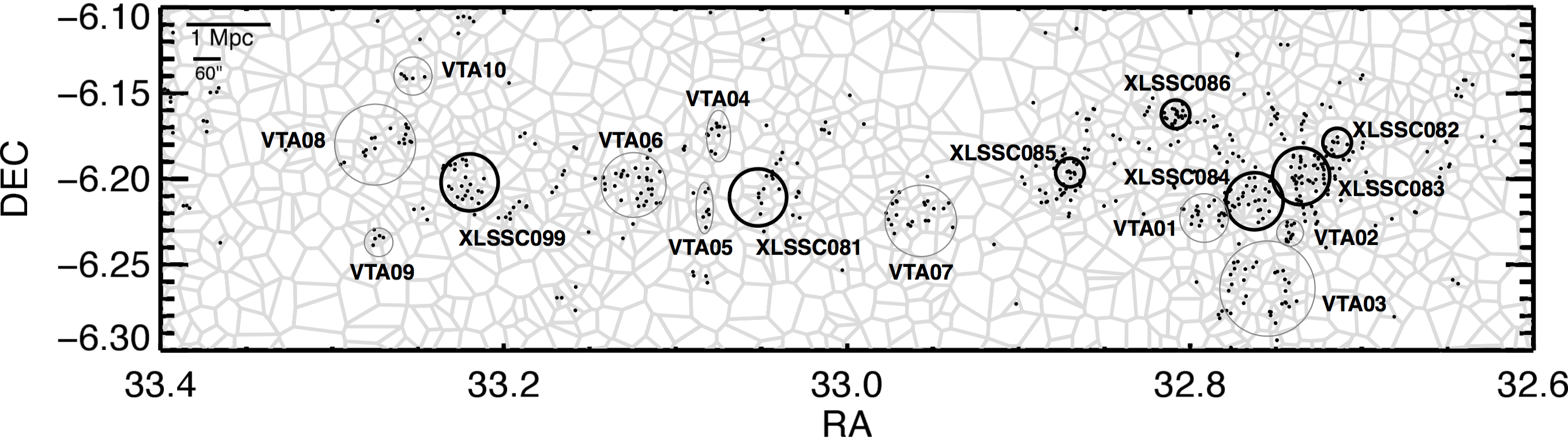}
		\protect\caption{Voronoi tessellation analysis of the area around the X-ray identified cluster structure at $0.35<z_{\mathrm{phot}}<0.50$
			(lines). Galaxies associated with overdense structures identified via our
			VTA analysis (see text for details) are shown as black dots. A projected scale of 1~Mpc is indicated in the top left corner. The X-ray identified clusters are marked with thick black circles, and VTA identified overdense structures are marked with thin grey circles, and labelled in the plot. Galaxies within these regions were sampled to statistically estimate the properties of clusters and overdense structures. (See Table \ref{tab:clusterstable} for details.)}
		\label{fig:voronoi_all} 
	\end{figure*}

	The results of the VTA at $z_{\mathrm{phot}}=0.35-0.50$ are shown
	in Fig.~\ref{fig:voronoi_all} where we indicate only potential overdense structure galaxy members. The regions from which galaxies were sampled were chosen so as to best encompass the VTA galaxy member candidates located either within the X-ray contours or within the VTA-identified overdensity structures (making sure that each galaxy was counted only in one overdensity).  We show
	an area larger than encompassed by the X-ray emission of the
	\nXMMt identified X-ray clusters to quantify potential additional overdense structures
	that may be undetected in the X-rays. In addition to the clusters
	already identified in the X-rays (labelled in Fig.~\ref{fig:voronoi_all};
	see also Fig.~1 in \citet{pompei15}), we identify \nVTAt additional galaxy
	overdense structures, labelled VTA01 to VTA10 in the figure. The
	overdense structures seem to assemble into two agglomerations (eastern and western),
	with an extended potential overdense structure in between (VTA07 in \f{fig:voronoi_all})
	. We note, however, that the area between these two agglomerations
	contains an optically masked area (a vertical strip at ${\rm RA}\sim32\fdg91$;
	cf. Fig.~1 in \citet{pompei15}). Thus, no clustering information
	could be retrieved for this part of the sky, which could contain additional
	overdense structures connecting the two agglomerations. The VTA may be not effective in detecting highly elongated, filamentary overdense structures. Since the structures in the field studied here are relatively close to each other, we do not expect this feature to influence our results significantly. Our VTA analysis based on photometric redshifts clearly identifies the clusters, independently detected via extended X-ray emission. This assures that the VTA yields robust group/cluster candidates, as further verified in the next section. 
	\subsubsection{Verification of the VTA-identified overdensities}
\begin{figure}[!]
		\includegraphics[width=0.9\linewidth]{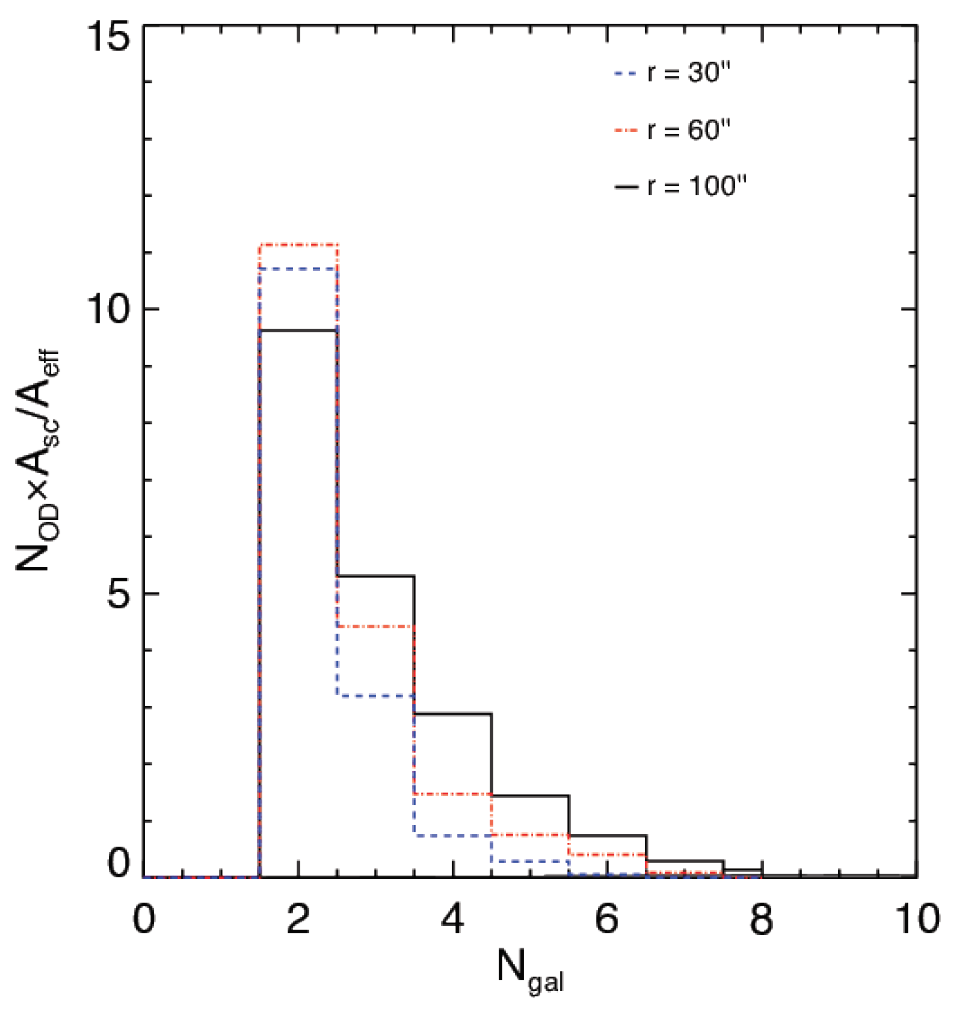}
		\caption{Expected number of circular overdense regions detected due to random density fluctuation as a function of the number of galaxies in these regions. The estimates are provided for three different radii, as indicated in the legend.}
		\label{fig:VoronoiTest}
	\end{figure}

	We test the probability of finding an overdensity caused by random density fluctuations using our te mock catalogues of randomly distributed galaxies across an area that is $\sim10$ times larger than that occupied by the supercluster. By tiling the plane with non-overlapping circular regions of a fixed radius, we identify regions with $N=1,2,3,...$ simulated overdense galaxies. We perform this analysis for three different circular areas with radii $r=30\arcsec$, $60\arcsec$, and $100\arcsec$. The result is then normalised by the effective area to derive the expected number of false positive detections of overdense regions. In Fig.~\ref{fig:VoronoiTest} we show the expected number of false positive detections with N galaxies within $r=30\arcsec$, $60\arcsec$, and $100\arcsec$ in our supercluster area. Since our overdense regions contain five or more galaxies within an effective radius of $\gtrsim30\arcsec$, we expect to find less than one such overdense region that is the result of random density fluctuations in our field. This is in agreement with \citet{ramella01} who set the minimum number of closely separated galaxies to five, as the threshold for cluster detection, rendering the effect of Poisson fluctuation negligible.
	
\begin{figure}[H]
		\centering
		\includegraphics[width=0.9\linewidth]{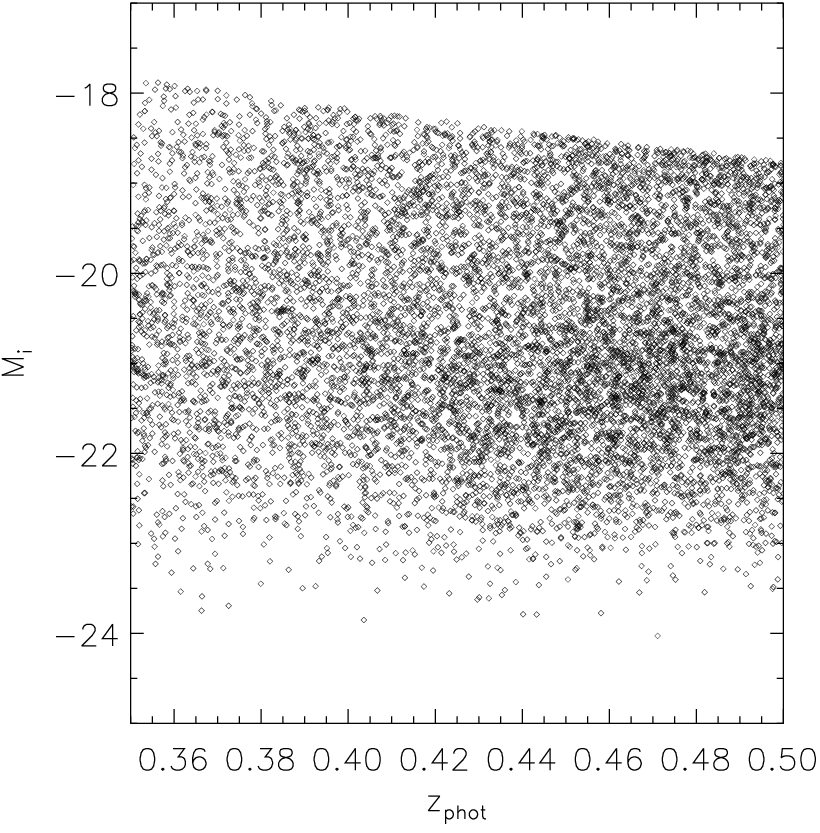}
		\includegraphics[width=0.9\linewidth]{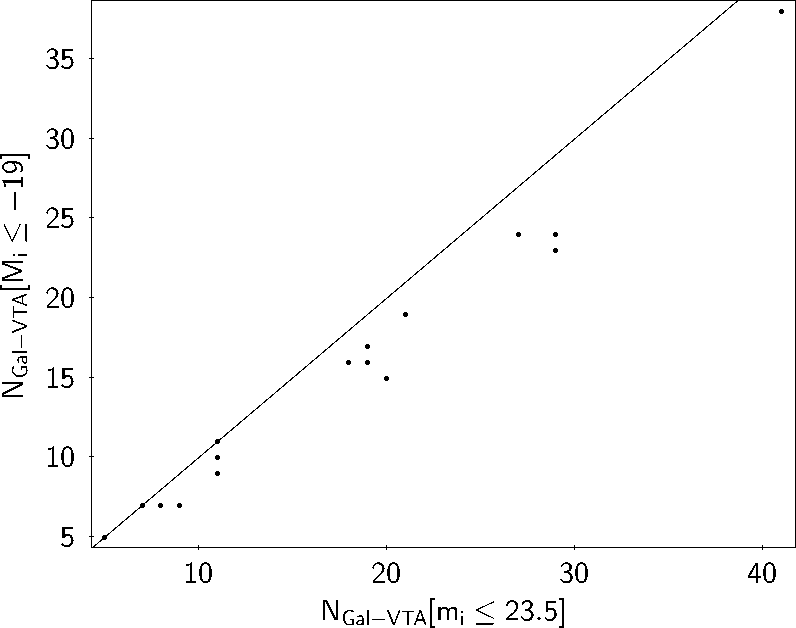}
		\caption{\textit{Top}: Absolute $i$~band magnitude as a function of photometric redshift. In the redshift range 0.35-0.5 a slight decrease can be seen. \textit{Bottom}: Comparison of the number of galaxies in each VTA overdense region with a cut in absolute $i$-band magnitude ($M_i\leq-19$), and a cut in observed $i$-band magnitude applied in this paper ($m_i\leq23.5$).}
		\label{fig:malmquist}
	\end{figure}

	To assess the significance of Malmquist bias on the detection of overdense structures we calculated absolute magnitudes for galaxies in the studied redshift range of 0.35~-~0.5. These absolute magnitudes are shown as a function of redshift in the upper panel of Fig.~\ref{fig:malmquist}. We note a drop in the absolute magnitude limit from $M_{i}\sim-18$ at $z=0.35$ to $M_{i}\sim-19$ at $z=0.5$. Cutting the optical catalogue at $M_{i}=-19$ instead of $m_{i}=23.5$, we find the same overdense regions, although with a slightly lower number of galaxies in each of them (lower panel in Fig.~\ref{fig:malmquist}). Thus, we conclude that Malmquist bias does not affect the results of the overdensity detections presented here.
	Finally, we performed an a posteriori verification of the validity of choosing the redshift range $0.35~<~z~<~0.5$. In the first step, we concatenated the CFHTLS photometric redshift and the CFHTLS photometric catalogue for an area $\sim10$ times the size of the supercluster. Using this catalogue we performed the VTA (as described in Sec.~\ref{subsub:vta}) for various $dz=0.15$ wide redshift ranges centred at $z=0.225~+~k~\times~0.1$, where $k=0,1,2,3,4,5$. In each redshift range the number of overdense galaxies within the $z=0.35-0.5$ identified overdensity regions were then identified.
	We find that all 17 overdense regions peak in the chosen redshift bin ($0.35-0.5$), suggesting this is the optimal range to describe all the overdense regions. However, to accurately report the mean redshift of each overdense region and the corresponding standard deviation we analyse each region within its own extent in redshift. We consider the region extended in redshift if the number of galaxies in redshift bins adjacent to the peak is within the Poissonian counting error of the peak value. We then use the extended redshift range to determine the mean redshift of overdense galaxies in each VTA region.
	Redshifts used to calculate the $\bar{z}_{phot}$ and its standard deviation for all overdense regions (including XLSSC clusters) are sampled from the range $0.35-0.5$, except for VTA02, VTA09, and VTA10. These overdense regions show a broader distribution in redshift, as determined by the criterion described above. Specifically, VTA02 was extended to $0.25<z<0.5$, VTA09 to $0.35<z<0.6$, and VTA10 to $0.25<z<0.5$.
	The mean photometric redshifts and the corresponding standard deviations of our VTA-identified
	overdense structure member candidates are listed in Table \ref{tab:clusterstable}.
	They are in agreement with the spectroscopic redshifts of all of the brightest
	cluster galaxies determined in \citet{pompei15}, also listed in Table~\ref{tab:clusterstable}.
	Spectroscopic redshift confirmation of the XLSSC~081-086 was performed on spectra of at least three galaxies (up to five) per cluster \citep{pompei15}. However, the spectroscopic redshift of the XLSSC~099 was determined from the spectrum of only one galaxy in the cluster (the brightest) using a spectrum from the GAMA\footnote{http://www.gama-survey.org/} survey. The median redshift error is $\sim$33km/s for the $z_{spec}$ of the GAMA objects \citep{baldry14}, rendering the $z_{spec}$ error negligible. The photometric redshift of this bright galaxy ($i$ = 17.88) is estimated to be $z_{phot}$ = 0.3615, and is thus consistent with its spectroscopic redshift, since $\sigma_{\Delta{z}/(1+z_{s})}\sim0.030-0.032$ for sources with $i<21.5$. Our analysis suggests that the mean overdensity redshift is slightly shifted relative to the redshift of this galaxy.
	
	\subsubsection{Galaxy colour properties}
\begin{figure}[!]
		\includegraphics[scale = 0.5]{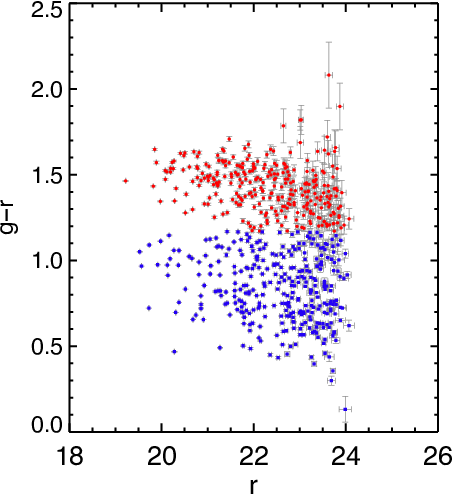} 
		\protect\caption{Colour-magnitude diagram of the sources in overdense structures within the XXL-N supercluster
			identified via the Voronoi tessellation analysis (see~Fig.~\ref{fig:voronoi_all}). Sources are classified as red (i.e. early-type) if their $g-r\ge1.17$, and blue (i.e. late-type) if their $g-r<1.17$.}
		\label{fig:colour-mag-all} 
	\end{figure}

\begin{figure*}[!]
		\includegraphics[width = 1\linewidth]{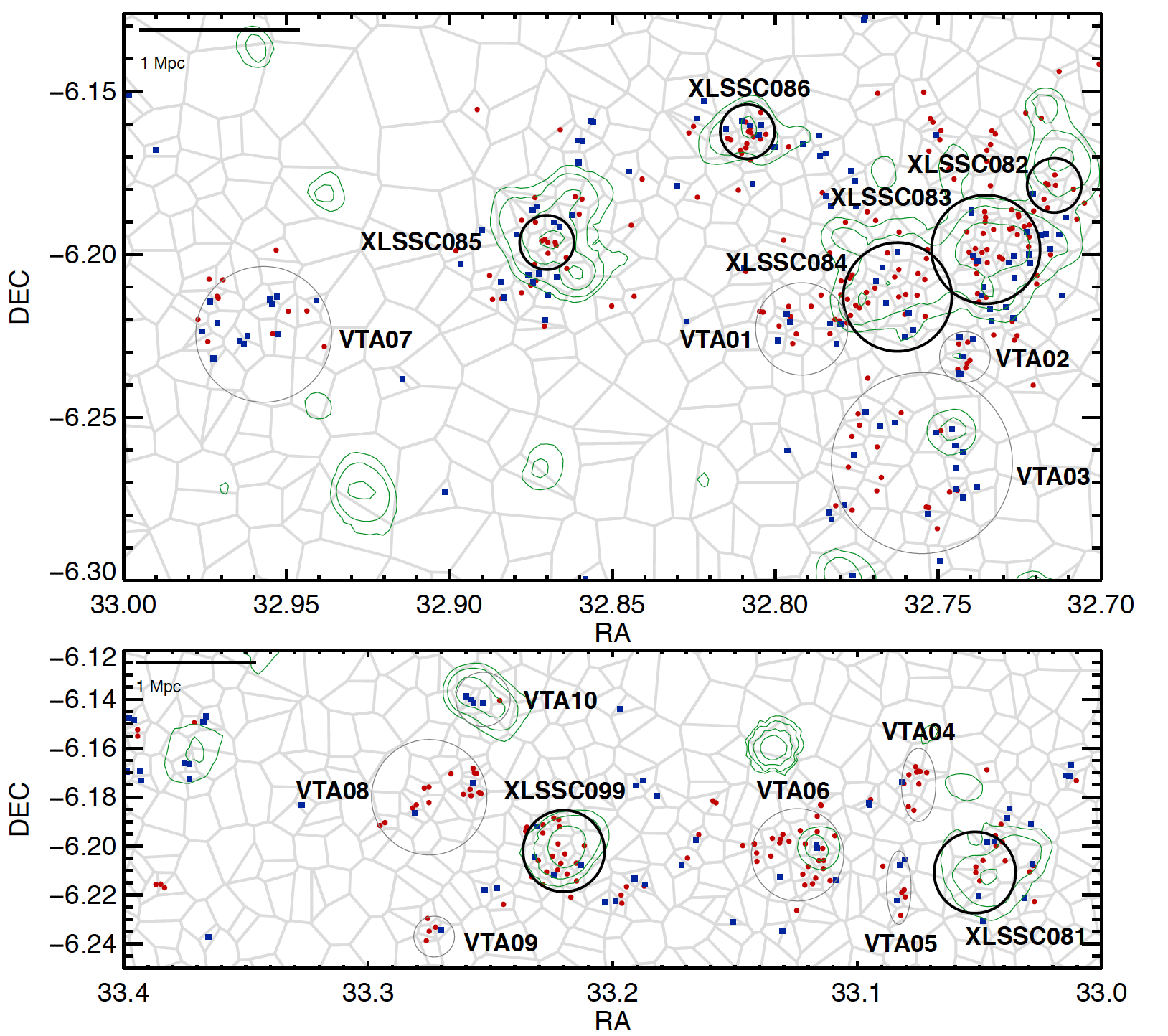}
		\protect\caption{Voronoi tessellation analysis of the western ($top$) and eastern ($bottom$) region of the supercluster
			at $0.35<z<0.50$ (grey lines). Potential cluster galaxy members,
			identified via our VTA analysis (see text for details), are separated
			into red ($g-r\ge1.17$) and blue ($g-r<1.17$) galaxies. These are shown
			by red filled circles and blue filled squares, respectively.
			The levels of the X-ray contours are shown in green at 5.28, 7.02, 12.54 and 30 cts~s$^{-1}$~deg~$^{-2}$ in the [0.5-2.5]~keV range. A scale of 1~Mpc is indicated in the top left corner. The X-ray identified clusters are marked with thick black circles, and VTA identified overdense structures are marked with thin grey circles, and are labelled in the plot. Galaxies within these regions were sampled to statistically estimate the properties of clusters and overdense structures (see text and Table~\ref{tab:clusterstable} for details).}
		\label{fig:voronoi_colour} 
	\end{figure*}

	We separate blue from red galaxies using a threshold of $g-r=1.17$. This value corresponds to the inflection point of the double-Gaussian function best fit to the $g-r$ histogram. The $g-r$~versus~$r$ colour-magnitude diagram for the VTA-identified
	potential supercluster galaxy members is shown in Fig.~\ref{fig:colour-mag-all}.
	In Fig.~\ref{fig:voronoi_colour} we
	show the spatial distribution of the red and blue potential supercluster
	galaxy members, overlaid with X-ray contours. In virialised clusters,
	red galaxies occupy the centre, while the blue galaxies occupy the
	outskirts of the cluster. As can be seen from the top and bottom panels of Fig.~\ref{fig:voronoi_colour},
	this does not appear to be the case here in all identified overdense structures. This will be further discussed in Sect.~\ref{sec:discussion}.
	
	\begin{table*}[!]
		\centering \protect\caption{List of VTA-detected cluster/group candidates. For clusters detected in
			the X-rays the X-ray ID (XLSSC) is adopted. }
		\tabcolsep=0.17cm
		\renewcommand{\footnoterule}{}
		
		\begin{tabular}{c c c c c c c c c c c c }
			
			\hline\hline 
			ID  & RA & Dec & $\bar{z}_{\textrm{{\scriptsize phot}}}$  & $\sigma_{\textrm{{\scriptsize phot}}}$ & $\bar{z}_{\textrm{{\scriptsize spec}}}$ & $L_{\textrm{\scriptsize 300}\textrm{kpc}}^{\scriptsize XXL}$ & $kT_{\scriptsize 300~\textrm{kpc}}$ &  $r_{500}$ & $M_{500, MT}$ \\
			& {\tiny {[}h m s{]}} & {\tiny {[}d m s{]}} & & & & {\tiny {[}$\times10^{42}$~erg~s$^{-1}${]}} & {\tiny {[}keV{]}} & {\tiny {[}kpc{]}} & {\tiny {[}$\times10^{13}$\msol{]}}\\
			(1)&(2)&(3)&(4)&(5)&(6)&(7)&(8)&(9)&(10)\\ 
			\hline 
			XLSSC~081 & 02 12 12\tablefootmark{*} & -06 12 39\tablefootmark{*} & 0.42  & 0.04  & 0.428\tablefootmark{*} &  14.9\tablefootmark{*}	 & 1.7\tablefootmark{*}  & 549\tablefootmark{*}& 7\tablefootmark{*}  \\
			XLSSC~082 & 02 10 51\tablefootmark{*} & -06 10 44\tablefootmark{*} & 0.42  & 0.04  & 0.424\tablefootmark{*} &  17.1\tablefootmark{*}	 & 3.9\tablefootmark{*}  & 878\tablefootmark{*}& 29\tablefootmark{*} \\
			XLSSC~083 & 02 10 56\tablefootmark{*} & -06 11 54\tablefootmark{*} & 0.42  & 0.03  & 0.430\tablefootmark{*} &  31.3\tablefootmark{*} & 4.8\tablefootmark{*}  & 990\tablefootmark{*}& 41\tablefootmark{*}  \\
			XLSSC~084 & 02 11 03\tablefootmark{*} & -06 12 47\tablefootmark{*} & 0.42  & 0.04  & 0.430\tablefootmark{*} &  13.8\tablefootmark{*} & 4.5\tablefootmark{*}  & 955\tablefootmark{*}& 37\tablefootmark{*}  \\
			XLSSC~085 & 02 11 29\tablefootmark{*} & -06 11 47\tablefootmark{*} & 0.43  & 0.03  & 0.428\tablefootmark{*} &  28.3\tablefootmark{*} & 4.8\tablefootmark{*}  & 985\tablefootmark{*}& 41\tablefootmark{*}  \\
			XLSSC~086 & 02 11 14\tablefootmark{*} & -06 09 44\tablefootmark{*} & 0.41  & 0.04  & 0.424\tablefootmark{*} &  11.2\tablefootmark{*}	 & 2.6\tablefootmark{*}  & 698\tablefootmark{*}& 15\tablefootmark{*}  \\
			XLSSC~099 & 02 12 53                  & -06 12 07                  & 0.43  & 0.04  & 0.3911& 14.4    				 & 5.1   			     & 1038     		   & 47  \\
			VTA01    & 02 11 10                  & -06 13 22                  & 0.40  & 0.04  & ...   & <7.3    	      & <1.8    		 & <560.6     		   & <7.7 \\ 
			VTA02    & 02 10 58                  & -06 13 53                  & 0.39  & 0.05  & ...   & <9.3    	      & <1.9    		 & <590.2     		   & <9.0 \\
			VTA03    & 02 11 01                  & -06 15 50                  & 0.43  & 0.04  & ...   & <5.9    	      & <1.6    		 & <535.2     		   & <6.7 \\
			VTA04    & 02 12 18                  & -06 10 30                  & 0.41  & 0.03  & ...   & <9.1    	      & <1.9    		 & <586.9     		   & <8.8 \\
			VTA05    & 02 12 20                  & -06 12 01                  & 0.42  & 0.05  & ...   & <9.9    	      & <2.0    		 & <597.1     		   & <9.3 \\
			VTA06	 & 02 12 30                  & -06 12 13                  & 0.43  & 0.03  & ...   & <11.7     				 & <2.1    			 & <618.9      		   & <10.3 \\
			VTA07    & 02 11 50                  & -06 13 28                  & 0.43  & 0.03  & ...   & <7.5    				 & <1.8    			 & <563.3     		   & <7.8 \\
			VTA08    & 02 12 01                  & -06 09 36                  & 0.43  & 0.04  & ...   & <8.83  					 & <2.04    		 & <606.6 		   & <9.7\\
			VTA09    & 02 13 06                  & -06 14 13                  & 0.47  & 0.06  & ...   & <8.8    				 & <1.9    				 & <582.7     		   & <8.6 \\  
			VTA10    & 02 13 01                  & -06 08 24                  & 0.39  & 0.06  & ...   & <11.9    				 & <2.30    				 & <649.7     		   & <11.9 \\ 
			\hline 
		\end{tabular}\label{tab:clusterstable}\\
		\tablefoot{Cluster candidates with unavailable spectroscopic redshift are marked by ellipses (...). Central right ascension and declination of regions shown in Fig.~\ref{fig:voronoi_all}, Fig.~\ref{fig:voronoi_colour} and Fig.~\ref{fig:mfmapsc} are given in column (2) and column (3). For the galaxies satisfying the VTA overdensity threshold and potentially belonging to the given structure as suggested by the circular or elliptical areas in Fig.~\ref{fig:voronoi_all}, Fig.~\ref{fig:voronoi_colour} and Fig.~\ref{fig:mfmapsc}, the mean photometric redshift and its standard deviation are given in columns (4) and (5), respectively. The spectroscopic redshift is given in column (6) (where available). Where available, X-ray luminosity in units of erg~s$^{-1}$, temperature in keV, $r_{500}$ in kpc and $M_{500, MT}$ in units of solar mass (M$_{\astrosun}$) are given in columns (7), (8), (9), and (10), respectively (as described in \citet{pompei15}; \citet{giles15}). $M_{500, MT}$ for XLSSC clusters has been derived in \citet{lieu15} using $M_{WL}~-~T$ relation. For X-ray non-detected VTA structures, $3\sigma$ upper limits are given (see Sect.~\ref{sec:discussion} for details).\\
			\tablefoottext{*}{Taken from \citet{pompei15}.}\\}
	\end{table*}
	
	\subsection{Radio properties}
	
	\label{sub:radioproperties}
	
	Our VLA-XXL-N mosaic covers only the western part of the supercluster
	(XLSSC~081-086, VTA01-03, and VTA07). The X-ray cluster XLSSC~081
	is on the very edge of the mosaic and encompasses one source detected in VLA-XXL-N radio observations (VLA-XXL ID = 154), but without an optical counterpart. VTA04-09 and XLSSC~099 are outside the field-of-view
	of our 3~GHz VLA-XXL-N mosaic. Thus, for the eastern part of the supercluster we search
	for radio sources associated with CFHTLS galaxies at $0.35<z_{\mathrm{phot}}<0.50$
	using the FIRST and NVSS 1.4~GHz survey catalogues (\citealt{becker95};
	\citealt{condon98}). We find no radio sources or radio galaxies
	associated with galaxies at $0.35<z_{\mathrm{phot}}<0.50$ in the
	area of interest.
	
\begin{figure*}
		\includegraphics[width = 1\linewidth]{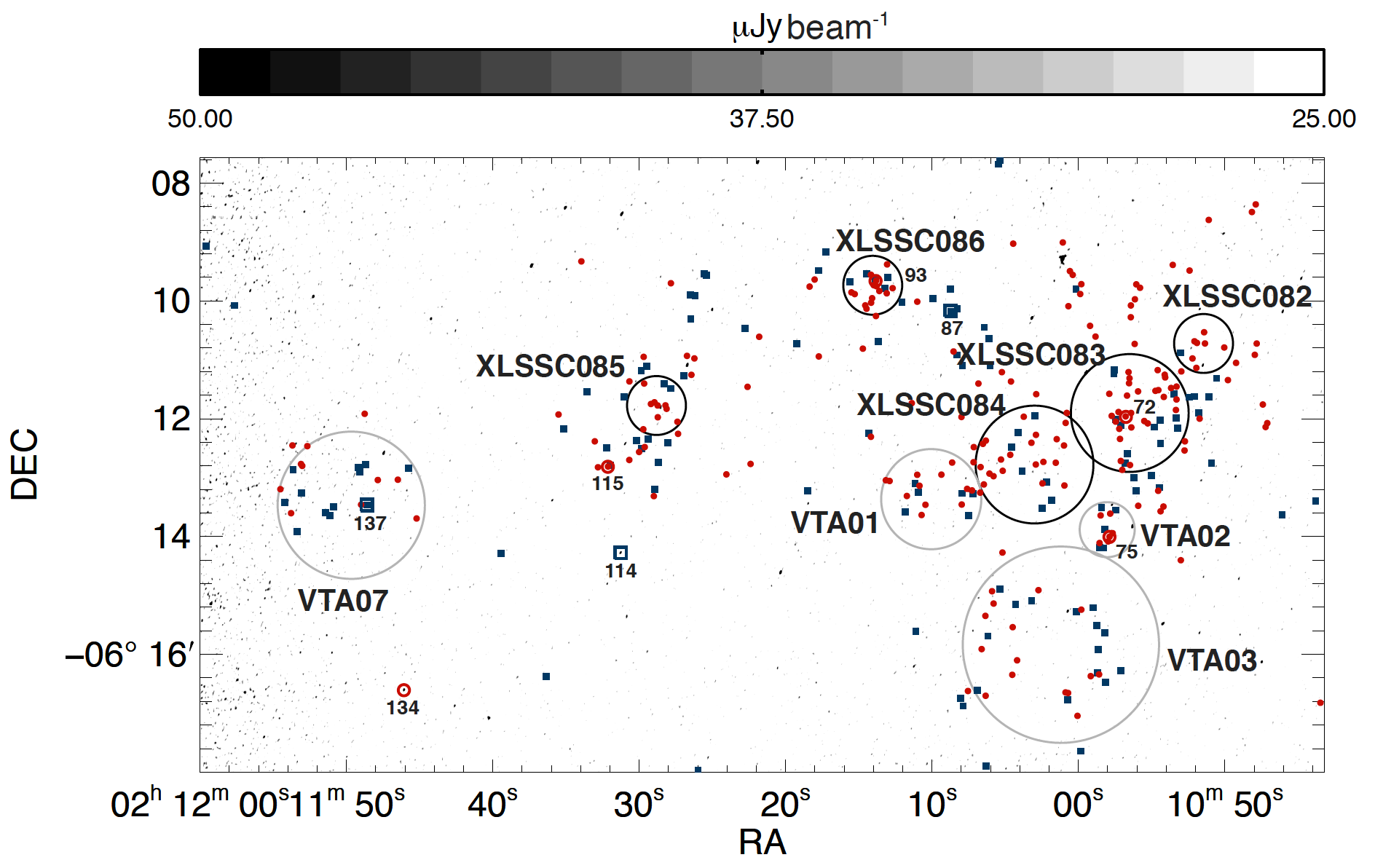} 
		\protect\caption{Area of VLA - XXL-N mosaic containing the supercluster. `Overdense'
			early-type galaxies are shown as red dots and 
			late-type galaxies as blue squares. Galaxies are classified as red
			if their $g-r\ge1.17$, and blue if $g-r<1.17$ (see Fig. \ref{fig:colour-mag-all}).
			Open red circles (blue squares) mark potential red (blue) supercluster
			member galaxies detected as 3~GHz radio sources (the 3~GHz VLA ID
			is indicated next to the radio sources). The X-ray identified clusters are marked with thick black circles, and VTA identified overdense structures are marked with thin grey circles, and labelled in the plot. Galaxies within these regions were sampled to statistically estimate the properties of clusters and overdense structures (see Table \ref{tab:clusterstable} for details).}
		\label{fig:mfmapsc} 
	\end{figure*}

\begin{figure*}
		\includegraphics[bb=70 0 637 789,scale=0.21]{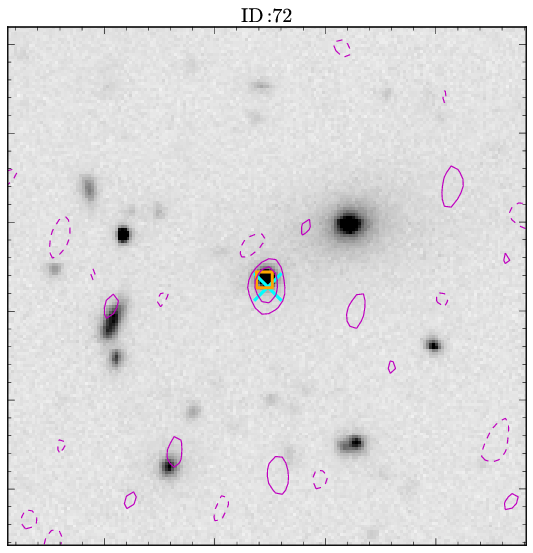}
		\includegraphics[bb=15 0 637 589,scale=0.21]{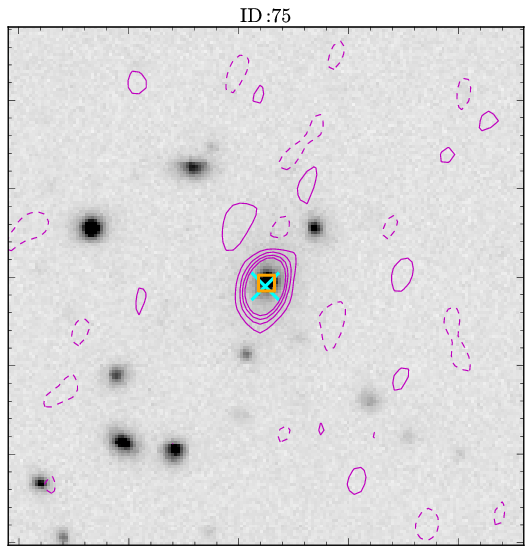}
		\includegraphics[bb=15 0 637 589,scale=0.21]{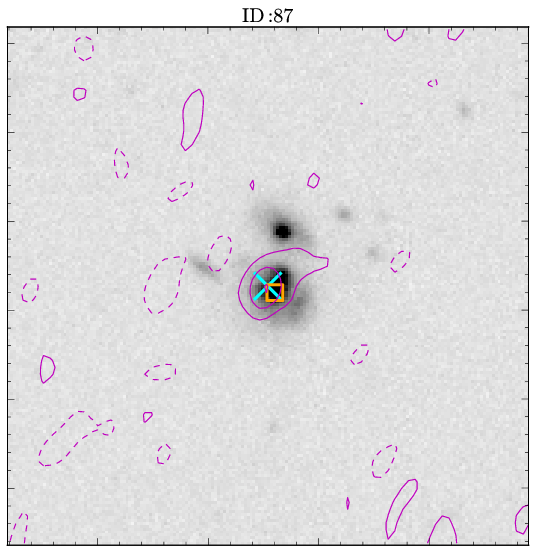}
		\includegraphics[bb=15 0 637 589,scale=0.21]{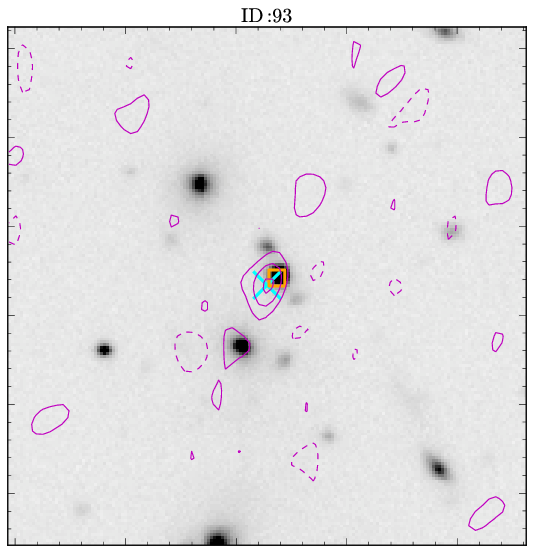}\\
		\includegraphics[bb=70 0 637 589,scale=0.21]{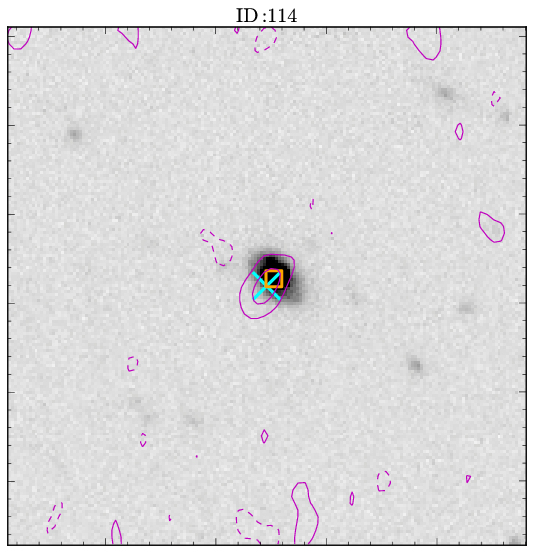}
		\includegraphics[bb=15 0 637 589,scale=0.21]{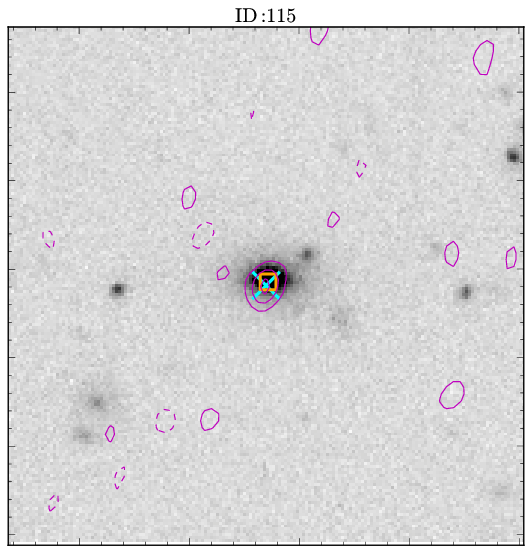}
		\includegraphics[bb=15 0 637 589,scale=0.21]{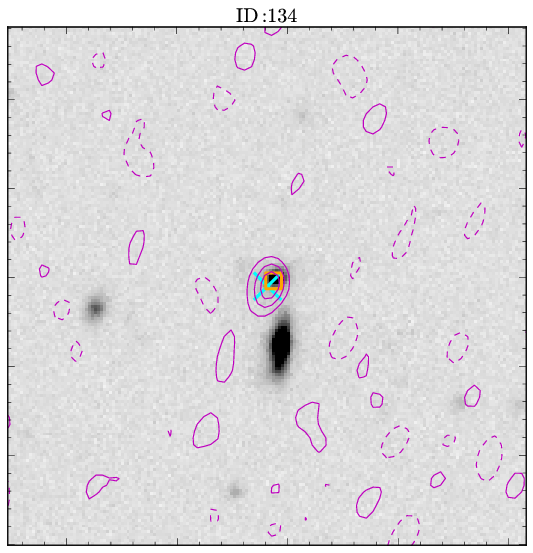}
		\includegraphics[bb=15 0 637 589,scale=0.21]{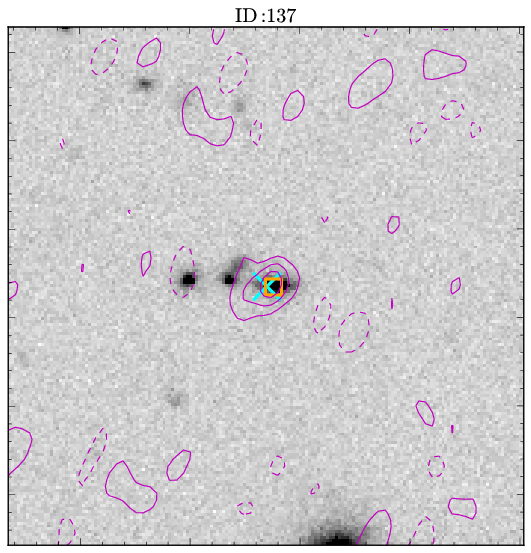}
		\protect\caption{Optical images (greyscale; $35\arcsec$ on the side) of potential
			supercluster member galaxies also detected at 3~GHz within the VLA-XXL
			data, overlaid with magenta radio contours. The contours start at 2$\sigma$
			($1\sigma=10.8~\mu$Jy) and increase in steps of 2$\sigma$ for all
			sources except source 75 where they increase in steps of 4$\sigma$.
			Dashed negative contours are shown at the same levels as the positive. Cyan crosses show the position of the of the VLA radio sources, orange squares represent the position of the optical counterpart. The VLA-XXL-N source IDs are shown above each panel. North is up, east is left.}
		
		\label{fig:stamps} 
	\end{figure*}

	In Fig. \ref{fig:mfmapsc} we show our 3~GHz radio mosaic of the
	area hosting the western part of the supercluster. In this area we
	find \nRadioInCluster 3~GHz radio sources with matched CFHTLS photometric redshifts between
	0.35 and 0.50. We show zoomed-in images of the host galaxies in the
	optical with radio contours overlaid in Fig.~\ref{fig:stamps}. No large radio galaxies are found within and around the X-ray identified cluster structure.
	
	The properties of the radio sources are listed in Table \ref{tab:radio}. Four out of the
	\nRadioInCluster sources are hosted by red galaxies, one has a colour consistent with blue galaxies, and two have host galaxy
	colours consistent with the green valley (upper part of the blue cloud; see \f{fig:colour-mag-all} ). A visual inspection of the
	host galaxies shows that the colours are, as expected, linked to the
	morphologies of the galaxies (e.g. \citealt{strateva01}). The red
	galaxies are elliptical, while the blue galaxies show spiral features.
	The 3~GHz luminosity densities of the sources (assuming a spectral index
	of $-0.8$) are in the range of $3\times10^{22}-2\times10^{23}$~\wh.
	Three out of the eight radio sources can be associated with the
	central regions of X-ray clusters or VTA identified overdense structures (VLA-XXL ID = 072 in XLSSC~083,
	VLA-XXL ID = 075 in VTA02 and VLA-XXL ID = 093 in XLSSC~086). The host galaxies
	of these are among the four brightest galaxies in the cluster/overdense structure. In particular,
	the host galaxy of VLA-XXL ID = 075 in the overdense structure VTA02 is the optically brightest
	galaxy in this structure.
	
	\begin{table*}
		\centering \protect\caption{Radio sources with optical counterparts that have redshifts $0.35<z_{\textrm {phot}}<0.50$ within
			the area containing the studied overdense structures.}
		
		\begin{tabular}{c c c c c c c c}
			\hline\hline 
			ID  & RA  & Dec  & $S_{\textrm{{\scriptsize peak}}}$  & $z_{\mathrm{phot}}$ & \lum  & $g-r$  & overdense structure ID\\
			& {[}h m s{]}  & {[}d m s{]}  & {[}$\mu$Jy~beam$^{-1}${]}  &  & {[}$\times10^{22}$ W/Hz{]}  &  & \\
			(1)&(2)&(3)&(4)&(5)&(6)&(7)&(8)\\ 
			\hline 
			114 & 02 11 31.27  & -06 14 16.84  & 59.5 $\pm$ 9.5   & 0.41 & 3.2 $\pm$ 0.5  & 1.09  & ... \\
			75  & 02 10 57.88  & -06 14 00.62  & 287.0 $\pm$ 10.8 & 0.41 & 16.1 $\pm$ 0.6 & 1.56  & VTA02\\
			115 & 02 11 32.15  & -06 12 49.11  & 74.5 $\pm$ 8.7   & 0.40 & 3.9 $\pm$ 0.4  & 1.62  & XLSSC~085 \\
			72  & 02 10 56.76  & -06 11 58.37  & 72.2 $\pm$ 9.7   & 0.41 & 4.0 $\pm$ 0.5  & 1.59  & XLSSC~083 \\
			87  & 02 11 08.73  & -06 10 10.38  & 66.6 $\pm$ 10.5  & 0.49 & 5.5 $\pm$ 0.8  & 1.02  & XLSSC~086 \\
			93  & 02 11 13.86  & -06 09 39.96  & 74.9 $\pm$ 10.8  & 0.40 & 4.0 $\pm$ 0.5  & 1.28  & XLSSC~086 \\
			137 & 02 11 48.62  & -06 13 27.91  & 80.6 $\pm$ 11.7  & 0.42 & 4.6 $\pm$ 0.6  & 0.71  & VTA07 \\
			134 & 02 11 46.09  & -06 16 36.59  & 91.7 $\pm$ 12.0  & 0.50 & 7.9 $\pm$ 1.0  & 1.50  & ... \\
			\hline 
		\end{tabular}\label{tab:radio} 
		\tablefoot{Sources unassociated with a cluster/group candidate are marked by ellipsis. Source identifiers from the VLA-XXL-N 3~GHz radio catalogue are given in column (1), column (2) gives the right ascension (RA) of the source in hours, minutes, and seconds, column (3) gives the declination (Dec) of the source in degrees, arcminutes and arcseconds. Peak surface brightness is given in column (4) in units of $\mu$Jy~beam$^{-1}$. Photometric redshifts of optical counterparts of radio sources are given in column (5). Radio luminosities at 3~GHz are given in column (6) in units of W/Hz, and are divided by $10^{22}$. Column (7) provides $g-r$ optical colour. If a source is associated with an overdense structure, column (8) gives the ID of the source.}
	\end{table*}
	
	\section{Discussion}
	
	\label{sec:discussion}
	
	\subsection{Constraining total masses of the overdensitites}
	
	In \citet{pompei15} the six westernmost
	X-ray detected clusters in the XLSSC-e supercluster have been spectroscopically confirmed in the region of interest to lie at $z=0.42-0.43$
	within an extent of $\sim 0^\circ\llap{.}35 \times 0^\circ\llap{.}1$ 
	on the sky (or a physical extent of $\sim10\times2.9$~Mpc$^{2}$ at $z=0.43$).
	Using the XXL X-ray data they have also computed the virial masses
	within the radius encompassing 500 times the critical density of
	the six clusters to be in the range of M$_{\mathrm{500}}=(1-3)\times10^{14}$~\msol. This yields a total mass of all six of the westernmost X-ray detected clusters of $M_{\mathrm{500}}\sim10^{15}$~\msol 
	(see Table 1 in \citet{pompei15}; see also Table 3 in \citet{pacaud15}). 
	
	Additionally, ten VTA identified overdense structures have not been identified as X-ray clusters, presumably because their X-ray emission falls below the X-ray cluster detection threshold. Nevertheless, for these overdense structures we can put upper limits on their X-ray properties as follows. The upper limits on VTA overdense structures were calculated using upper limits on the count
	rates in a 300~kpc aperture centred at the middle of the optical structures listed in Table~\ref{tab:clusterstable}.
	The method consists of the Bayesian approach for calculating X-ray aperture photometry described in detail in
	J.~Willis et al. (in prep.). It computes values and bounds for the intensity of the source (S) using counts and 
	exposure data obtained in source and background apertures. It calculates the background-marginalised posterior probability distribution function of the source intensity (S), assuming Poisson likelihoods for the detected number of source counts and background counts in the given exposure time. The mode of this PDF is determined, and the lower and upper bounds 
	of the confidence region are determined by summing values of the PDF alternately above and below the mode until 
	the desired confidence level is attained. When the mode is at S=0 or 
	the calculation for the lower bound reaches the value S=0, only the upper confidence bound is evaluated,  
	which is considered an upper limit.
	With this upper limit on the count rate in a 300~kpc aperture we used the mass-temperature relation
	of \citet{lieu15} (hereafter \citet{lieu15}) and the luminosity-temperature relation of \citet{giles15}, (hereafter \citet{giles15}) to calculate upper limits on the mass of the VTA overdense structures. Starting with an initial estimate of the temperature of the object, we use the $M-T$ relation to evaluate the mass and the radius corresponding to the overdense structure of 500 times the critical density. We integrate the given count rate to compute the X-ray luminosity up to $r_{500}$, and re-estimate the temperature predicted from the corresponding $L-T$ relation. We re-iterate until we converge on the value of the temperature. The calculations for the upper limits are given in Table~\ref{tab:clusterstable}. For VTA10 we masked two AGN point sources leaving only 48\% of the area of
	the 300~kpc aperture for the calculation of the upper limit on the flux. As expected, the upper limits on total masses estimated for our VTA-identified overdensities without X-ray detections are consistent with the structures being poor clusters or groups (e.g. \citealt{vajgel14}). A spectroscopic velocity dispersion of the structures would allow the classification of these structures (e.g. \citealt{dehghan14}). However, no such spectroscopic data is currently available.
	
	\subsection{Morphology and composition of the supercluster }
	
	As is evident from Fig. \ref{fig:voronoi_colour}, the X-ray emission of the clusters XLSSC~082, 083, and 084 overlap into a morphologically
	elongated ensemble in the NW-SE direction. Our VTA analysis reveals
	multiple overdense structures of optical galaxies in the same field, three
	of which lie at the same redshift but are not detected in the X-ray
	(VTA01 - VTA03). Overdense structure VTA01 is located toward the SE from
	the ensemble revealed in the X-rays, and appears to be a distinct
	structure. The overdense structures VTA02 and VTA03 are located toward the
	south of the ensemble. The non-detection of the VTA-identified overdense structures
	in the X-ray data sets the upper limit to the combined total virial mass of these three overdense structures to 2.3$\times10^{14}$~\msol (see Table~\ref{tab:clusterstable}.). 
	
	The spatial distribution of red
	and blue galaxies in the X-ray ensemble (XLSSC~082, XLSSC~083, XLSSC~084) and VTA01-03 groups appears
	non-structured, i.e. there is no clear red-blue galaxy separation
	with increasing distance from the group centre, which suggests an unvirialised
	state of the groups. If this is the case, then the calculated virial masses may overestimate the real masses of these cluster candidates.
	
	We find a 3~GHz radio source (VLA-XXL ID = 072) associated with the third
	brightest galaxy within the XLSSC~083 clusters close to the peak of
	the X-ray emission. Its monochromatic 3~GHz radio luminosity density is $(4.05~\pm~0.55)\times10^{22}$~\wh
	, and it is associated with a red-sequence galaxy ($g-r=1.6$). This suggests
	that the radio synchrotron emission likely arises from a weak AGN within the galaxy, often found at the bottom of the
	gravitational potential wells of galaxy clusters (e.g. \citealt{smolcic08, smolcic10}).
	
	The X-ray, radio and optical data combined, thus suggest that the
	bottom of the gravitational potential well is within the X-ray peak, and
	that the six clusters and overdense structures identified in this region (XLSSC~082, 083, 084 and VTA01, 02, 03) are likely in the process of merging and
	forming a larger structure. Based on the $M_{\mathrm{500}}$ values
	calculated in \citet{pompei15}, and using the upper mass limits given in Table~\ref{tab:clusterstable}, we can estimate that the upper limit to the total mass of the western structure after adding $M_{\mathrm{500}}$ is $M_{\mathrm{500}}\lesssim1.3\times10^{15}$~\msol.
	
	The clusters XLSSC~085 and 086 are at a projected distance of $8\arcmin$ and
	$4\farcm8$, respectively, from the XLSSC~083 cluster. This corresponds
	to a distance of $\sim2.7$ and $\sim\rm 1.6~Mpc$, respectively at $z=0.43$.
	Our VTA identifies overdense structures associated with both clusters. The red-blue
	galaxy distribution in XLSSC~086 appears non-structured, and both structures
	show subclumping and elongated overdense structure features. It is interesting
	that the brightest galaxy (in the $r$ band) in XLSSC~086 is associated with
	a red galaxy toward the south of the X-ray emission, while the next
	three brightest galaxies (one of them hosting the radio source VLA-XXL ID = 093)
	are associated with the X-ray centroid. The E-W
	elongation of the X-ray emission of this cluster, supported by a similar
	elongation of the distribution of galaxies associated with VTA-identified overdense structures, suggests the existence of two substructures that are in the process of merging.
	
	The X-ray detected cluster XLSSC~081 lies $\sim0.35$~deg away from
	XLSSC~083 (see Fig.~\ref{fig:voronoi_colour}), while the cluster XLSSC~099 is located further to the east, $\sim0.43$~deg from XLSSC~083. Furthermore, the VTA associates an elongated overdense structure of galaxies with this cluster. In addition
	to this cluster, our VTA has revealed seven more overdense structures
	toward the east of this cluster (VTA04-10), and a potential
	overdense structure toward the west (VTA07). All overdense structures identified
	in this region seem to have a non-segregated distribution of red and blue galaxies. This again suggests
	a dynamically young state for the structures. The X-ray emission coincident with VTA10 originates from two background X-ray sources, spectroscopically found to lie at $z_{spec}\sim0.6$ and $z_{spec}\sim2.1$ by SDSS and reported in \citet{koulouridis15}.
	
	In summary, the VTA resulted in the identification of \nTott
	overdense structures at $z_{\mathrm{phot}}=0.43$, \nXMMt of which are associated
	with X-ray detected clusters, while the remaining 10 structures are likely groups. The full structure extends over $0^\circ\llap{.}6 \times 0^\circ\llap{.}2$
	on the sky, which corresponds to a size of $\sim12\times4$~Mpc$^{2}$ at $z=0.43$. Two main galaxy cluster agglomerations
	are located toward the east (XLSSC~081, XLSSC~099, VTA04-10) and west (XLSSC~082-086,
	VTA01-03) of this structure. The identified overdense structures seem unvirialised
	and dynamically young, suggesting that these structures are perhaps
	in the process of forming a larger structure. To shed further light on the dynamical state of the overdensity structures, in the next section we investigate  the radio properties of the western cluster/group agglomeration covered by our 3~GHz observations.
	
	\subsection{Radio properties of the western cluster/group agglomeration}
	
	Previous studies used radio source counts and luminosity functions to investigate enhancement or supression of radio emission. Several authors have found that in the pre-merging stages the suppression of AGN activity has not yet reached the cluster core \citep{venturi97, venturi01} where radio AGN preferably reside \citep{dressler80}. A different situation is found in clusters that have already likely undergone the first core-to-core encounter and are now accreting smaller groups. In such cases, an excess in the number of blue galaxies is found at the expected positions of the shock fronts \citep{miller05, johnston08} as the merging process has had enough time to influence the galaxies' optical and radio properties. This implies that cluster mergers in later stages lower the probability of a galaxy developing an AGN \citep{venturi00, venturi01}, but have also been shown to enhance the number of low-power radio galaxies driven by star formation \citep{miller05, johnston08}. In this sense, as can be seen in Fig.~\ref{fig:mfmapsc}, the clusters XLSSC~082-084 are close to each other, and we find that the core of XLSSC~083 seems to be preserved. Thus it appears that the agglomeration of XLSSC~082-084 has not yet undergone the first core-to-core encounter. The cluster XLSSC~086 also contains a radio source within a red host galaxy at its centre, and we note the increase in number of blue galaxies between XLSSC~086 and the large cluster structure south-west of it, possibly indicating a shock front present in this area. This picture seems to be consistent with the merger process discussed by \citet{venturi01} - most massive structures are the first to undergo a merger (in this case XLSSC~082-084), and they later attract satellite structures of lower mass (in this case XLSSC~086, and possibly VTA02-03).
	
\begin{figure}[!hp]
		\centering
		\includegraphics[width=1\linewidth]{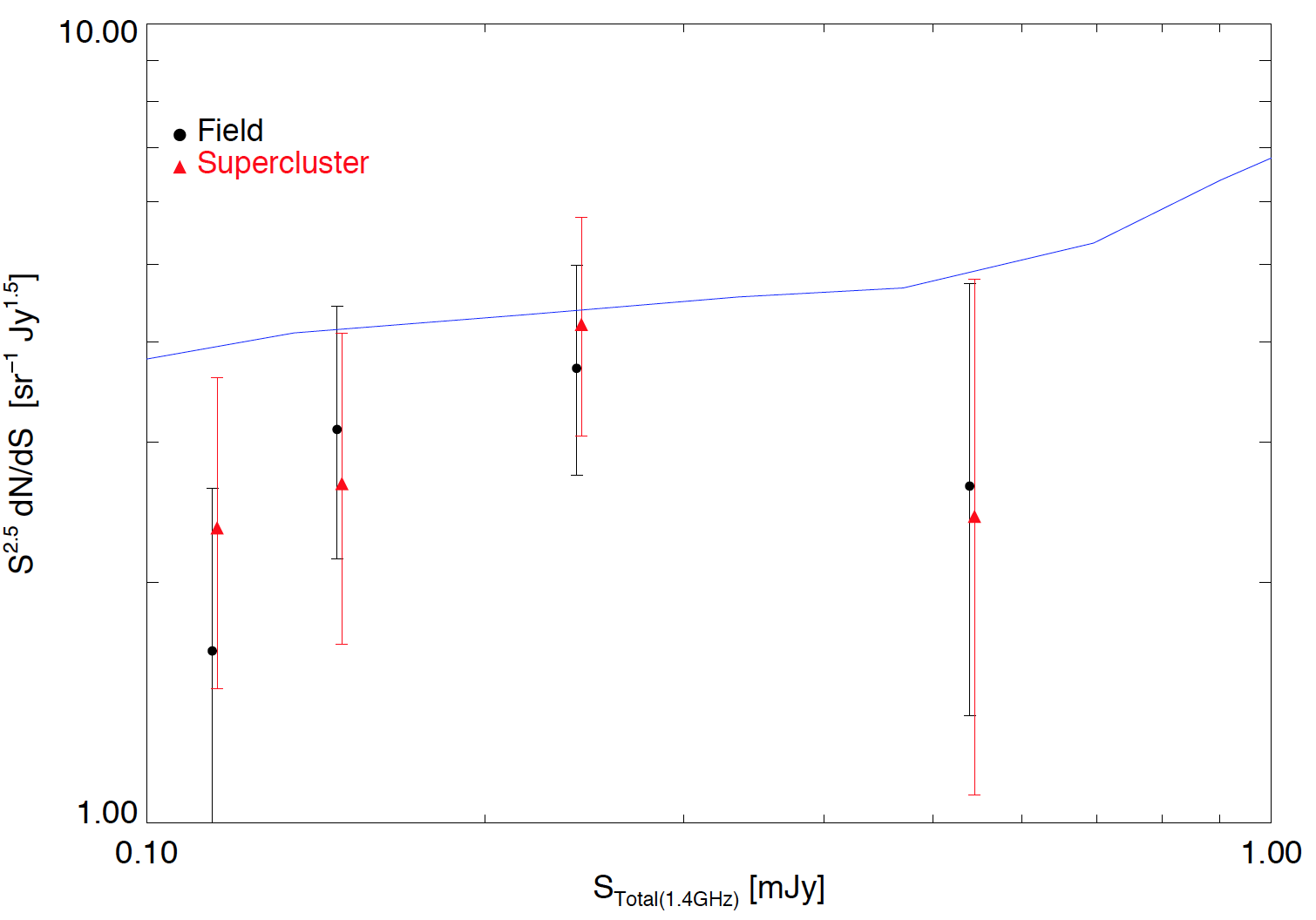}
		\caption{1.4~GHz radio source counts for the supercluster area (red triangles), and the field with good rms excluding the supercluster area (black). Red dots are slightly offset from the bin centres for better visualisation. Errors are Poissonian, calculated using approximate algebraic expressions from \citet{gehrels86}. Radio source counts from \citet{condon12} are shown by the blue line. The flux densities at 1.4~GHz were calculated from the 3~GHz fluxes assuming a spectral index of -0.7 for consistency with \citet{condon12}.}
		\label{fig:counts}
	\end{figure}

	To further investigate this, in Fig.~\ref{fig:counts} we show the Euclidean normalised radio source counts for the supercluster area, and the field with good rms excluding the supercluster area, as well as the general field source counts taken from \citet{condon12}. Given the low number of radio sources in our field the supercluster counts are consistent with those in the field. Our results suggest the radio counts within the supercluster may be slightly enhanced (suppressed) at low (high) fluxes relative to the field. This is similar to the suppresion of radio AGNs and an enhancement of low-power (likely star-forming) radio galaxies found in Shapely Supercluster \citep{venturi00, venturi02, miller05} and the Horologium-Reticulum Supercluster \citep{johnston08} at $z\sim$0.07. However, this is statistically not significant given the large error bars, and it prevents us from drawing further conclusions.
	
	Our radio data probe $L_{3~\mathrm{GHz}}\gtrsim4\times10^{21}~\mathrm{W/Hz}$. This limit is very close to the division between star-forming and AGN galaxies, thus in the range where volume densities of star-forming and AGN galaxies are comparable. Converting this limit to star formation rate (SFR) using the \citet{bell03} relation yields SFR$\gtrsim30$\Msol/year. Thus, our data are not deep enough to probe a potential enhancement of star-forming galaxies within the supercluster for $\lesssim30$\msolyr. However, they are sensitive enough to detect powerful radio AGNs usually hosted by red galaxies (e.g. \citealt{best06}, \citealt{smolcic08}), if present within the supercluster.
	
\begin{figure}
		\centering
		\includegraphics[bb = 0 100 595 742, angle = 90, width=1\linewidth]{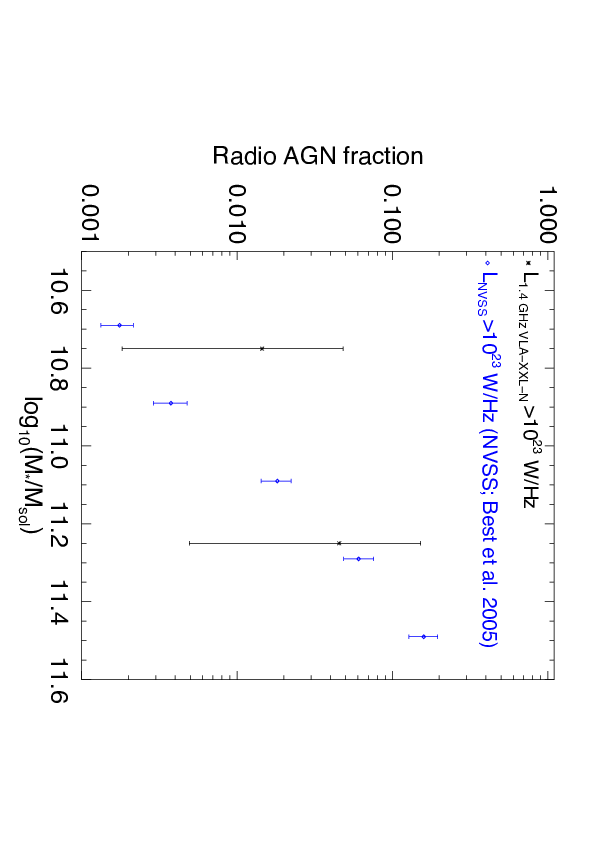}
		\caption{The fraction of radio detected red galaxies (radio AGN) as a function of the stellar mass for the region shown in the upper panel of Fig.~\ref{fig:voronoi_colour}. Black stars show the fractions estimated in this paper for galaxies with $L_{1.4~\mathrm{GHz}}>10^{23}~\mathrm{W/Hz}$. The errors shown on the black points are calculated from the equations given in \citet{gehrels86}. Blue diamonds show the fractions obtained for the NVSS sample by \citet{best05}.}
		\label{fig:StellarMass}
	\end{figure}

	In Fig.~\ref{fig:StellarMass} we show the fraction of red galaxies detected in the radio (radio AGNs, as mentioned above) with $L_{1.4~\mathrm{GHz}}>10^{23}~\mathrm{W/Hz}$ as a function of stellar mass for the region shown in the upper panel of Fig.~\ref{fig:voronoi_colour}. The fraction was calculated by taking the ratio of the number of red ($g-r\geq1.17$) galaxies from the supercluster area detected in the radio (see Tab.~\ref{tab:radio}) with $L_{1.4~\mathrm{GHz}}>10^{23}~\mathrm{W/Hz}$, and the number of optical red galaxies in the same field. The radio luminosity threshold was chosen to validate the comparison with the red galaxy fraction of NVSS-detected sources hosted by red galaxies from \citet{best05}, and also shown in Fig.~\ref{fig:StellarMass}.
	
	The fractions derived here for the general field derived by \citet{best05} agree well for masses above $log(\rm{M_*}/$\Msol$)\sim11$, however, for $log(\rm{M_*}/$\Msol$)\lesssim10.8$ the fraction of radio AGNs is slightly higher than expected, although still within the errors.
	
	If the excess in the fraction of radio detected galaxies of lower stellar mass ($log(\rm{M_*}/$\Msol$)\sim10.8$) were real, it could be explained by the physical mechanisms in the cluster which favour the AGN activity in red galaxies. \citet{best05} find that richer environments preferably host radio AGN. For galaxies with stellar masses of $\sim10^{11}$\msol they estimate a cooling rate of the order of magnitude of $\sim1$\msolyr, which is consistent with a low-power radio AGN. This indicates that the cooling of gas in the hot atmosphere of the host galaxy is a plausible candidate for the fuel source of radio AGN in the cluster structure studied here. This leads to the possibility that this cluster structure is rich in cooling gas, and/or that the cooling gas is likely cospatial with the lower-mass galaxies.
	
	In summary, by analysing the properties of radio sources within the structure, we conclude that the clusters and group candidates in the western agglomeration are likely in the pre-merger state, before the first core-to-core encounter. An increase in the surface density of blue galaxies between the XLSSC~086 and the XLSSC~082-084 agglomeration could indicate a formation of a shock front. Since the XLSSC~082-084 agglomeration is the most massive one in the area, it is likely that in the future the three most massive clusters will be the first to go radio silent, followed by the smaller satellite structures XLSSC~086 and VTA01-03.
	
	\section{Summary}
	
	\label{sec:summary}
	
	We have presented observations of a subfield of the XXL-N 25~deg$^{2}$
	field with the VLA data at 3~GHz (10~cm). The final radio mosaic
	has a resolution of $3\farcs2~\times~1\farcs9$ and encompasses
	an area of $41\arcmin\times41\arcmin$ with $rms\lesssim20$~$\mu$Jy~beam$^{-1}$.
	The rms in the central $15\arcmin\times15\arcmin$ area is $10.8~\mu$Jy~beam$^{-1}$.
	From the mosaic we have extracted 155 sources with S/N$>6$, 8 of
	which are large, multicomponent sources. Of the 155 sources,
	we find optical counterparts for 123 sources within the CFHTLS W1
	catalogue.
	
	We also analysed the first supercluster discovered in the XXL project,
	partially located within the radio mosaic. Applying the Voronoi tessellation
	analysis using photometric redshifts from the CFHTLS survey have
	identified \nTott overdense structures at $z_{\mathrm{phot}}=0.35-0.50$,
	\nXMMt of which are detected in the \textit{XMM-Newton} XXL data. The structure
	is extended over $0^\circ\llap{.}6 \times 0^\circ\llap{.}2$ on the sky, which corresponds
	to a size of $\sim12\times4$~Mpc$^{2}$ at $z=0.43$.
	No large radio galaxies are present within the overdense structures, and
	we associate eight (S/N$>$7) radio sources with potential supercluster
	members. We find that the spatial distribution of the red and blue
	potential group member galaxies, selected by their observed
	$g-r$ colours, suggests that the groups are not virialised, but are
	dynamically young, which is consistent with the hierarchical structure growth expected
	in a $\Lambda$CDM universe. Further spectroscopic follow-up is required
	for a more detailed analysis of the dynamical state of the structure.
	\begin{acknowledgements}
		XXL is an international project based around an XMM Very Large Programme surveying two 25~deg$^2$ extragalactic fields at a depth of $\sim5~\times~10-15$~erg~cm$^{-2}$~s$^{-1}$ in the [0.5-2]~keV band for point-like sources. The XXL website is \texttt{http://irfu.cea.fr/xxl}. Multiband information and spectroscopic follow-up of the X-ray sources are obtained through a number of survey programmes, summarised at \texttt{http://xxlmultiwave.pbworks.com/}.
		\textit{Funding} Saclay group: long-term support by the Centre National d'Etudes Spatiales (CNES). F.P. : support from BMBF DLR grant 50 OR 1117 and support from the DfG Transregio Programme TR33. Zgal group: European Union's Seventh Framework program under grant agreement 333654 (CIG, 'AGN feedback'; N.B., V.S.) and grant agreement 337595 (ERC Starting Grant, 'CoSMass'; V.S., J.D., O.Mi., M. N.). Italian and French groups: support from the International Programme for Scientific Cooperation CNRS-INAF PICS 2012. O.Me. acknowledges for the financial support provided by the European Union Seventh Framework Programme (FP7 2007-2013), grant agreement N\textsuperscript{\underline{o}} 291823 Marie Curie FP7-PEOPLE-2011-COFUND (The new International Fellowship Mobility Programme for Experienced Researchers in Croatia - NEWFELPRO, project "AGN environs in XXL", Grant Agreement \#83). We thank the referee for a very useful report.
		Authors of this paper acknowledge Olivier Ilbert's valuable input.
		This work is based on observations obtained with XMM-Newton, an ESA science mission
		with instruments and contributions directly funded by ESA Member States and NASA. This work is also based on observations obtained with MegaPrime/MegaCam, a
		joint project of CFHT and CEA/DAPNIA, at the Canada-France-Hawaii
		Telescope (CFHT) which is operated by the National Research Council
		(NRC) of Canada, the Institut National des Sciences de l'Univers of
		the Centre National de la Recherche Scientifique (CNRS) of France, and
		the University of Hawaii. This research used the facilities of the
		Canadian Astronomy Data Centre operated by the National Research
		Council of Canada with the support of the Canadian Space Agency.
		CFHTLenS data processing was made possible thanks to significant
		computing support from the NSERC Research Tools and Instruments grant
		program.
	\end{acknowledgements}
	
	\bibliographystyle{aa}
	
	\bibliography{nb_bibtex}

\begin{thebibliography}{54}
\expandafter\ifx\csname natexlab\endcsname\relax\def\natexlab#1{#1}\fi

\bibitem[{{Arnouts} {et~al.}(1999){Arnouts}, {Cristiani}, {Moscardini},
  {Matarrese}, {Lucchin}, {Fontana}, \& {Giallongo}}]{arnouts99}
{Arnouts}, S., {Cristiani}, S., {Moscardini}, L., {et~al.} 1999, \mnras, 310,
  540

\bibitem[{{Baldry} {et~al.}(2014){Baldry}, {Alpaslan}, {Bauer},
  {Bland-Hawthorn}, {Brough}, {Cluver}, {Croom}, {Davies}, {Driver},
  {Gunawardhana}, {Holwerda}, {Hopkins}, {Kelvin}, {Liske},
  {L{\'o}pez-S{\'a}nchez}, {Loveday}, {Norberg}, {Peacock}, {Robotham}, \&
  {Taylor}}]{baldry14}
{Baldry}, I.~K., {Alpaslan}, M., {Bauer}, A.~E., {et~al.} 2014, \mnras, 441,
  2440

\bibitem[{{Becker} {et~al.}(1995){Becker}, {White}, \& {Helfand}}]{becker95}
{Becker}, R.~H., {White}, R.~L., \& {Helfand}, D.~J. 1995, \apj, 450, 559

\bibitem[{{Bell}(2003)}]{bell03}
{Bell}, E.~F. 2003, \apj, 586, 794

\bibitem[{{Best} {et~al.}(2006){Best}, {Kaiser}, {Heckman}, \&
  {Kauffmann}}]{best06}
{Best}, P.~N., {Kaiser}, C.~R., {Heckman}, T.~M., \& {Kauffmann}, G. 2006,
  \mnras, 368, L67

\bibitem[{{Best} {et~al.}(2005){Best}, {Kauffmann}, {Heckman}, {Brinchmann},
  {Charlot}, {Ivezi{\'c}}, \& {White}}]{best05}
{Best}, P.~N., {Kauffmann}, G., {Heckman}, T.~M., {et~al.} 2005, \mnras, 362,
  25

\bibitem[{{Bondi} {et~al.}(2008){Bondi}, {Ciliegi}, {Schinnerer}, {Smol{\v
  c}i{\'c}}, {Jahnke}, {Carilli}, \& {Zamorani}}]{bondi08}
{Bondi}, M., {Ciliegi}, P., {Schinnerer}, E., {et~al.} 2008, \apj, 681, 1129

\bibitem[{{Bourke} {et~al.}(2014){Bourke}, {Mooley}, \& {Hallinan}}]{bourke14}
{Bourke}, S., {Mooley}, K., \& {Hallinan}, G. 2014, in Astronomical Society of
  the Pacific Conference Series, Vol. 485, Astronomical Data Analysis Software
  and Systems XXIII, ed. N.~{Manset} \& P.~{Forshay}, 367

\bibitem[{{Chabrier}(2003)}]{chabrier03}
{Chabrier}, G. 2003, \pasp, 115, 763

\bibitem[{{Condon} {et~al.}(2012){Condon}, {Cotton}, {Fomalont}, {Kellermann},
  {Miller}, {Perley}, {Scott}, {Vernstrom}, \& {Wall}}]{condon12}
{Condon}, J.~J., {Cotton}, W.~D., {Fomalont}, E.~B., {et~al.} 2012, \apj, 758,
  23

\bibitem[{{Condon} {et~al.}(1998){Condon}, {Cotton}, {Greisen}, {Yin},
  {Perley}, {Taylor}, \& {Broderick}}]{condon98}
{Condon}, J.~J., {Cotton}, W.~D., {Greisen}, E.~W., {et~al.} 1998, \aj, 115,
  1693

\bibitem[{{Coupon} {et~al.}(2009){Coupon}, {Ilbert}, {Kilbinger}, {McCracken},
  {Mellier}, {Arnouts}, {Bertin}, {Hudelot}, {Schultheis}, {Le F{\`e}vre}, {Le
  Brun}, {Guzzo}, {Bardelli}, {Zucca}, {Bolzonella}, {Garilli}, {Zamorani},
  {Zanichelli}, {Tresse}, \& {Aussel}}]{coupon09}
{Coupon}, J., {Ilbert}, O., {Kilbinger}, M., {et~al.} 2009, \aap, 500, 981

\bibitem[{{Dehghan} \& {Johnston-Hollitt}(2014)}]{dehghan14}
{Dehghan}, S. \& {Johnston-Hollitt}, M. 2014, \aj, 147, 52

\bibitem[{{Dickinson} {et~al.}(2003){Dickinson}, {Giavalisco}, \& {GOODS
  Team}}]{dickinson03}
{Dickinson}, M., {Giavalisco}, M., \& {GOODS Team}. 2003, in The Mass of
  Galaxies at Low and High Redshift, ed. R.~{Bender} \& A.~{Renzini}, 324

\bibitem[{{Dressler}(1980)}]{dressler80}
{Dressler}, A. 1980, \apj, 236, 351

\bibitem[{{Driver} {et~al.}(2011){Driver}, {Hill}, {Kelvin}, {Robotham},
  {Liske}, {Norberg}, {Baldry}, {Bamford}, {Hopkins}, {Loveday}, {Peacock},
  {Andrae}, {Bland-Hawthorn}, {Brough}, {Brown}, {Cameron}, {Ching}, {Colless},
  {Conselice}, {Croom}, {Cross}, {de Propris}, {Dye}, {Drinkwater}, {Ellis},
  {Graham}, {Grootes}, {Gunawardhana}, {Jones}, {van Kampen}, {Maraston},
  {Nichol}, {Parkinson}, {Phillipps}, {Pimbblet}, {Popescu}, {Prescott},
  {Roseboom}, {Sadler}, {Sansom}, {Sharp}, {Smith}, {Taylor}, {Thomas},
  {Tuffs}, {Wijesinghe}, {Dunne}, {Frenk}, {Jarvis}, {Madore}, {Meyer},
  {Seibert}, {Staveley-Smith}, {Sutherland}, \& {Warren}}]{driver11}
{Driver}, S.~P., {Hill}, D.~T., {Kelvin}, L.~S., {et~al.} 2011, \mnras, 413,
  971

\bibitem[{{Driver} {et~al.}(2009){Driver}, {Norberg}, {Baldry}, {Bamford},
  {Hopkins}, {Liske}, {Loveday}, {Peacock}, {Hill}, {Kelvin}, {Robotham},
  {Cross}, {Parkinson}, {Prescott}, {Conselice}, {Dunne}, {Brough}, {Jones},
  {Sharp}, {van Kampen}, {Oliver}, {Roseboom}, {Bland-Hawthorn}, {Croom},
  {Ellis}, {Cameron}, {Cole}, {Frenk}, {Couch}, {Graham}, {Proctor}, {De
  Propris}, {Doyle}, {Edmondson}, {Nichol}, {Thomas}, {Eales}, {Jarvis},
  {Kuijken}, {Lahav}, {Madore}, {Seibert}, {Meyer}, {Staveley-Smith},
  {Phillipps}, {Popescu}, {Sansom}, {Sutherland}, {Tuffs}, \&
  {Warren}}]{driver09}
{Driver}, S.~P., {Norberg}, P., {Baldry}, I.~K., {et~al.} 2009, Astronomy and
  Geophysics, 50, 12

\bibitem[{{Durret} {et~al.}(2011){Durret}, {Adami}, {Cappi}, {Maurogordato},
  {M{\'a}rquez}, {Ilbert}, {Coupon}, {Arnouts}, {Benoist}, {Blaizot}, {Edorh},
  {Garilli}, {Guennou}, {Le Brun}, {Le F{\`e}vre}, {Mazure}, {McCracken},
  {Mellier}, {Mezrag}, {Slezak}, {Tresse}, \& {Ulmer}}]{durret11}
{Durret}, F., {Adami}, C., {Cappi}, A., {et~al.} 2011, \aap, 535, A65

\bibitem[{{Erben} {et~al.}(2013){Erben}, {Hildebrandt}, {Miller}, {van
  Waerbeke}, {Heymans}, {Hoekstra}, {Kitching}, {Mellier}, {Benjamin}, {Blake},
  {Bonnett}, {Cordes}, {Coupon}, {Fu}, {Gavazzi}, {Gillis}, {Grocutt}, {Gwyn},
  {Holhjem}, {Hudson}, {Kilbinger}, {Kuijken}, {Milkeraitis}, {Rowe},
  {Schrabback}, {Semboloni}, {Simon}, {Smit}, {Toader}, {Vafaei}, {van Uitert},
  \& {Velander}}]{erben13}
{Erben}, T., {Hildebrandt}, H., {Miller}, L., {et~al.} 2013, \mnras, 433, 2545

\bibitem[{{Gehrels}(1986)}]{gehrels86}
{Gehrels}, N. 1986, \apj, 303, 336

\bibitem[{{Giles} {et~al.}(2015){Giles}, {Maughan}, {Pacaud}, {Lieu}, {Clerc},
  {Pierre}, {Adami}, {Chiappetti}, {D{\'e}mocl{\'e}s}, {Ettori}, {Le
  F{\'e}vre}, {Ponman}, {Sadibekova}, {Smith}, {Willis}, \& {Ziparo}}]{giles15}
{Giles}, P.~A., {Maughan}, B.~J., {Pacaud}, F., {et~al.} 2015, ArXiv e-prints

\bibitem[{{Grogin} {et~al.}(2011){Grogin}, {Kocevski}, {Faber}, {Ferguson},
  {Koekemoer}, {Riess}, {Acquaviva}, {Alexander}, {Almaini}, {Ashby}, {Barden},
  {Bell}, {Bournaud}, {Brown}, {Caputi}, {Casertano}, {Cassata}, {Castellano},
  {Challis}, {Chary}, {Cheung}, {Cirasuolo}, {Conselice}, {Roshan Cooray},
  {Croton}, {Daddi}, {Dahlen}, {Dav{\'e}}, {de Mello}, {Dekel}, {Dickinson},
  {Dolch}, {Donley}, {Dunlop}, {Dutton}, {Elbaz}, {Fazio}, {Filippenko},
  {Finkelstein}, {Fontana}, {Gardner}, {Garnavich}, {Gawiser}, {Giavalisco},
  {Grazian}, {Guo}, {Hathi}, {H{\"a}ussler}, {Hopkins}, {Huang}, {Huang},
  {Jha}, {Kartaltepe}, {Kirshner}, {Koo}, {Lai}, {Lee}, {Li}, {Lotz}, {Lucas},
  {Madau}, {McCarthy}, {McGrath}, {McIntosh}, {McLure}, {Mobasher},
  {Moustakas}, {Mozena}, {Nandra}, {Newman}, {Niemi}, {Noeske}, {Papovich},
  {Pentericci}, {Pope}, {Primack}, {Rajan}, {Ravindranath}, {Reddy}, {Renzini},
  {Rix}, {Robaina}, {Rodney}, {Rosario}, {Rosati}, {Salimbeni}, {Scarlata},
  {Siana}, {Simard}, {Smidt}, {Somerville}, {Spinrad}, {Straughn}, {Strolger},
  {Telford}, {Teplitz}, {Trump}, {van der Wel}, {Villforth}, {Wechsler},
  {Weiner}, {Wiklind}, {Wild}, {Wilson}, {Wuyts}, {Yan}, \& {Yun}}]{grogin11}
{Grogin}, N.~A., {Kocevski}, D.~D., {Faber}, S.~M., {et~al.} 2011, \apjs, 197,
  35

\bibitem[{{Helfand} {et~al.}(2015){Helfand}, {White}, \& {Becker}}]{helfand15}
{Helfand}, D.~J., {White}, R.~L., \& {Becker}, R.~H. 2015, \apj, 801, 26

\bibitem[{{Heymans} {et~al.}(2012){Heymans}, {Van Waerbeke}, {Miller}, {Erben},
  {Hildebrandt}, {Hoekstra}, {Kitching}, {Mellier}, {Simon}, {Bonnett},
  {Coupon}, {Fu}, {Harnois D{\'e}raps}, {Hudson}, {Kilbinger}, {Kuijken},
  {Rowe}, {Schrabback}, {Semboloni}, {van Uitert}, {Vafaei}, \&
  {Velander}}]{heymans12}
{Heymans}, C., {Van Waerbeke}, L., {Miller}, L., {et~al.} 2012, \mnras, 427,
  146

\bibitem[{{Hildebrandt} {et~al.}(2012){Hildebrandt}, {Erben}, {Kuijken}, {van
  Waerbeke}, {Heymans}, {Coupon}, {Benjamin}, {Bonnett}, {Fu}, {Hoekstra},
  {Kitching}, {Mellier}, {Miller}, {Velander}, {Hudson}, {Rowe}, {Schrabback},
  {Semboloni}, \& {Ben{\'{\i}}tez}}]{hildebrandt12}
{Hildebrandt}, H., {Erben}, T., {Kuijken}, K., {et~al.} 2012, \mnras, 421, 2355

\bibitem[{{Hinshaw} {et~al.}(2013){Hinshaw}, {Larson}, {Komatsu}, {Spergel},
  {Bennett}, {Dunkley}, {Nolta}, {Halpern}, {Hill}, {Odegard}, {Page}, {Smith},
  {Weiland}, {Gold}, {Jarosik}, {Kogut}, {Limon}, {Meyer}, {Tucker}, {Wollack},
  \& {Wright}}]{hinshaw13}
{Hinshaw}, G., {Larson}, D., {Komatsu}, E., {et~al.} 2013, \apjs, 208, 19

\bibitem[{{Ilbert} {et~al.}(2006){Ilbert}, {Arnouts}, {McCracken},
  {Bolzonella}, {Bertin}, {Le F{\`e}vre}, {Mellier}, {Zamorani}, {Pell{\`o}},
  {Iovino}, {Tresse}, {Le Brun}, {Bottini}, {Garilli}, {Maccagni}, {Picat},
  {Scaramella}, {Scodeggio}, {Vettolani}, {Zanichelli}, {Adami}, {Bardelli},
  {Cappi}, {Charlot}, {Ciliegi}, {Contini}, {Cucciati}, {Foucaud}, {Franzetti},
  {Gavignaud}, {Guzzo}, {Marano}, {Marinoni}, {Mazure}, {Meneux}, {Merighi},
  {Paltani}, {Pollo}, {Pozzetti}, {Radovich}, {Zucca}, {Bondi}, {Bongiorno},
  {Busarello}, {de La Torre}, {Gregorini}, {Lamareille}, {Mathez}, {Merluzzi},
  {Ripepi}, {Rizzo}, \& {Vergani}}]{ilbert06}
{Ilbert}, O., {Arnouts}, S., {McCracken}, H.~J., {et~al.} 2006, \aap, 457, 841

\bibitem[{{Jeli{\'c}} {et~al.}(2012){Jeli{\'c}}, {Smol{\v c}i{\'c}},
  {Finoguenov}, {Tanaka}, {Civano}, {Schinnerer}, {Cappelluti}, \&
  {Koekemoer}}]{jelic12}
{Jeli{\'c}}, V., {Smol{\v c}i{\'c}}, V., {Finoguenov}, A., {et~al.} 2012,
  \mnras, 423, 2753

\bibitem[{{Johnston-Hollitt} {et~al.}(2008){Johnston-Hollitt}, {Sato}, {Gill},
  {Fleenor}, \& {Brick}}]{johnston08}
{Johnston-Hollitt}, M., {Sato}, M., {Gill}, J.~A., {Fleenor}, M.~C., \&
  {Brick}, A.-M. 2008, \mnras, 390, 289

\bibitem[{{Kimball} \& {Ivezi{\'c}}(2008)}]{kimball08}
{Kimball}, A.~E. \& {Ivezi{\'c}}, {\v Z}. 2008, \aj, 136, 684

\bibitem[{{Koekemoer} {et~al.}(2011){Koekemoer}, {Faber}, {Ferguson}, {Grogin},
  {Kocevski}, {Koo}, {Lai}, {Lotz}, {Lucas}, {McGrath}, {Ogaz}, {Rajan},
  {Riess}, {Rodney}, {Strolger}, {Casertano}, {Castellano}, {Dahlen},
  {Dickinson}, {Dolch}, {Fontana}, {Giavalisco}, {Grazian}, {Guo}, {Hathi},
  {Huang}, {van der Wel}, {Yan}, {Acquaviva}, {Alexander}, {Almaini}, {Ashby},
  {Barden}, {Bell}, {Bournaud}, {Brown}, {Caputi}, {Cassata}, {Challis},
  {Chary}, {Cheung}, {Cirasuolo}, {Conselice}, {Roshan Cooray}, {Croton},
  {Daddi}, {Dav{\'e}}, {de Mello}, {de Ravel}, {Dekel}, {Donley}, {Dunlop},
  {Dutton}, {Elbaz}, {Fazio}, {Filippenko}, {Finkelstein}, {Frazer}, {Gardner},
  {Garnavich}, {Gawiser}, {Gruetzbauch}, {Hartley}, {H{\"a}ussler},
  {Herrington}, {Hopkins}, {Huang}, {Jha}, {Johnson}, {Kartaltepe},
  {Khostovan}, {Kirshner}, {Lani}, {Lee}, {Li}, {Madau}, {McCarthy},
  {McIntosh}, {McLure}, {McPartland}, {Mobasher}, {Moreira}, {Mortlock},
  {Moustakas}, {Mozena}, {Nandra}, {Newman}, {Nielsen}, {Niemi}, {Noeske},
  {Papovich}, {Pentericci}, {Pope}, {Primack}, {Ravindranath}, {Reddy},
  {Renzini}, {Rix}, {Robaina}, {Rosario}, {Rosati}, {Salimbeni}, {Scarlata},
  {Siana}, {Simard}, {Smidt}, {Snyder}, {Somerville}, {Spinrad}, {Straughn},
  {Telford}, {Teplitz}, {Trump}, {Vargas}, {Villforth}, {Wagner}, {Wandro},
  {Wechsler}, {Weiner}, {Wiklind}, {Wild}, {Wilson}, {Wuyts}, \&
  {Yun}}]{koekemoer11}
{Koekemoer}, A.~M., {Faber}, S.~M., {Ferguson}, H.~C., {et~al.} 2011, \apjs,
  197, 36

\bibitem[{{Koulouridis} {et~al.}(2015){Koulouridis}, {Poggianti}, {Altieri},
  {Valchanov}, {Jaff{\'e}}, {Adami}, {Elyiv}, {Melnyk}, {Fotopoulou},
  {Gastaldello}, {Horellou}, {Pierre}, {Pacaud}, {Plionis}, {Sadibekova}, \&
  {Surdej}}]{koulouridis15}
{Koulouridis}, E., {Poggianti}, B., {Altieri}, B., {et~al.} 2015, ArXiv
  e-prints

\bibitem[{{Lieu} {et~al.}(2015){Lieu}, {Smith}, {Giles}, {Ziparo}, {Maughan},
  {D{\'e}mocl{\`e}s}, {Pacaud}, {Pierre}, {Adami}, {Bah{\'e}}, {Clerc},
  {Chiappetti}, {Eckert}, {Ettori}, {Lavoie}, {Le Fevre}, {McCarthy},
  {Kilbinger}, {Ponman}, {Sadibekova}, \& {Willis}}]{lieu15}
{Lieu}, M., {Smith}, G.~P., {Giles}, P.~A., {et~al.} 2015, ArXiv e-prints

\bibitem[{{Miller}(2005)}]{miller05}
{Miller}, N.~A. 2005, \aj, 130, 2541

\bibitem[{{Mooley} {et~al.}(2015){Mooley}, {Hallinan}, {Frail}, {Myers},
  {Kulkarni}, {Bourke}, \& {Horesh}}]{mooley15}
{Mooley}, K.~P., {Hallinan}, G., {Frail}, D.~A., {et~al.} 2015, in American
  Astronomical Society Meeting Abstracts, Vol. 225, American Astronomical
  Society Meeting Abstracts, 113.05

\bibitem[{{Novak} {et~al.}(2015){Novak}, {Smol{\v c}i{\'c}}, {Civano}, {Bondi},
  {Ciliegi}, {Wang}, {Loeb}, {Banfield}, {Bourke}, {Elvis}, {Hallinan},
  {Intema}, {Kl{\"o}ckner}, {Mooley}, \& {Navarrete}}]{novak15}
{Novak}, M., {Smol{\v c}i{\'c}}, V., {Civano}, F., {et~al.} 2015, \mnras, 447,
  1282

\bibitem[{{Oklop{\v c}i{\'c}} {et~al.}(2011){Oklop{\v c}i{\'c}}, {Smol{\v
  c}i{\'c}}, {Giodini}, {Zamorani}, {Bhatirzan}, {Schinnerer}, {Carilli},
  {Finoguenov}, {Lilly}, {Koekemoer}, \& {Scoville}}]{oklopcic11}
{Oklop{\v c}i{\'c}}, A., {Smol{\v c}i{\'c}}, V., {Giodini}, S., {et~al.} 2011,
  \memsai, 82, 161

\bibitem[{{Pacaud} {et~al.}(2015){Pacaud}, {Clerc}, {Giles}, {Adami},
  {Sadibekova}, {Pierre}, {Maughan}, {Lieu}, {Le F{\`e}vre}, {Alis}, {Altieri},
  {Ardila}, {Baldry}, {Benoist}, {Birkinshaw}, {Chiappetti},
  {D{\'e}mocl{\`e}s}, {Eckert}, {Evrard}, {Faccioli}, {Gastaldello}, {Guennou},
  {Horellou}, {Iovino}, {Koulouridis}, {Le Brun}, {Lidman}, {Liske},
  {Maurogordato}, {Menanteau}, {Owers}, {Poggianti}, {Pomar{\`e}de}, {Pompei},
  {Ponman}, {Rapetti}, {Reiprich}, {Smith}, {Tuffs}, {Valageas}, {Valtchanov},
  {Willis}, \& {Ziparo}}]{pacaud15}
{Pacaud}, F., {Clerc}, N., {Giles}, P.~A., {et~al.} 2015, ArXiv e-prints

\bibitem[{{Pierre} {et~al.}(2015){Pierre}, {Pacaud}, {Adami}, {Alis},
  {Altieri}, {Baran}, {Benoist}, {Birkinshaw}, {Bongiorno}, {Bremer}, {Brusa},
  {Butler}, {Ciliegi}, {Chiappetti}, {Clerc}, {Corasaniti}, {Coupon}, {De
  Breuck}, {Democles}, {Desai}, {Delhaize}, {Devriendt}, {Dubois}, {Eckert},
  {Elyiv}, {Ettori}, {Evrard}, {Faccioli}, {Farahi}, {Ferrari}, {Finet},
  {Fotopoulou}, {Fourmanoit}, {Gandhi}, {Gastaldello}, {Gastaud},
  {Georgantopoulos}, {Giles}, {Guennou}, {Guglielmo}, {Horellou}, {Husband},
  {Huynh}, {Iovino}, {Kilbinger}, {Koulouridis}, {Lavoie}, {Le Brun}, {Le
  Fevre}, {Lidman}, {Lieu}, {Lin}, {Mantz}, {Maughan}, {Maurogordato},
  {McCarthy}, {McGee}, {Melin}, {Melnyk}, {Menanteau}, {Novak}, {Paltani},
  {Plionis}, {Poggianti}, {Pomarede}, {Pompei}, {Ponman}, {Ramos-Ceja},
  {Ranalli}, {Rapetti}, {Raychaudury}, {Reiprich}, {Rottgering}, {Rozo},
  {Rykoff}, {Sadibekova}, {Santos}, {Sauvageot}, {Schimd}, {Sereno}, {Smith},
  {Smol{\v c}i{\'c}}, {Snowden}, {Spergel}, {Stanford}, {Surdej}, {Valageas},
  {Valotti}, {Valtchanov}, {Vignali}, {Willis}, \& {Ziparo}}]{pierre15}
{Pierre}, M., {Pacaud}, F., {Adami}, C., {et~al.} 2015, ArXiv e-prints

\bibitem[{{Pompei} {et~al.}(2015){Pompei}, {Adami}, {Eckert}, {Gastaldello},
  {Lavoie}, {Poggianti}, {Altieri}, {Alis}, {Baran}, {Benoist}, {Jaffe'},
  {Koulouridis}, {Maurogordato}, {Pacaud}, {Pierre}, {Sadibekova}, {Smolcic},
  \& {Valtchanov}}]{pompei15}
{Pompei}, E., {Adami}, C., {Eckert}, D., {et~al.} 2015, ArXiv e-prints

\bibitem[{{Ramella} {et~al.}(2001){Ramella}, {Boschin}, {Fadda}, \&
  {Nonino}}]{ramella01}
{Ramella}, M., {Boschin}, W., {Fadda}, D., \& {Nonino}, M. 2001, \aap, 368, 776

\bibitem[{{Rau} \& {Cornwell}(2011)}]{rau11}
{Rau}, U. \& {Cornwell}, T.~J. 2011, \aap, 532, A71

\bibitem[{{Scoville} {et~al.}(2007){Scoville}, {Aussel}, {Brusa}, {Capak},
  {Carollo}, {Elvis}, {Giavalisco}, {Guzzo}, {Hasinger}, {Impey}, {Kneib},
  {LeFevre}, {Lilly}, {Mobasher}, {Renzini}, {Rich}, {Sanders}, {Schinnerer},
  {Schminovich}, {Shopbell}, {Taniguchi}, \& {Tyson}}]{scoville07}
{Scoville}, N., {Aussel}, H., {Brusa}, M., {et~al.} 2007, \apjs, 172, 1

\bibitem[{{Smol{\v c}i{\'c}} {et~al.}(2007){Smol{\v c}i{\'c}}, {Schinnerer},
  {Finoguenov}, {Sakelliou}, {Carilli}, {Botzler}, {Brusa}, {Scoville},
  {Ajiki}, {Capak}, {Guzzo}, {Hasinger}, {Impey}, {Jahnke}, {Kartaltepe},
  {McCracken}, {Mobasher}, {Murayama}, {Sasaki}, {Shioya}, {Taniguchi}, \&
  {Trump}}]{smolcic07}
{Smol{\v c}i{\'c}}, V., {Schinnerer}, E., {Finoguenov}, A., {et~al.} 2007,
  \apjs, 172, 295

\bibitem[{{Smol{\v c}i{\'c}} {et~al.}(2008){Smol{\v c}i{\'c}}, {Schinnerer},
  {Scodeggio}, {Franzetti}, {Aussel}, {Bondi}, {Brusa}, {Carilli}, {Capak},
  {Charlot}, {Ciliegi}, {Ilbert}, {Ivezi{\'c}}, {Jahnke}, {McCracken},
  {Obri{\'c}}, {Salvato}, {Sanders}, {Scoville}, {Trump}, {Tremonti}, {Tasca},
  {Walcher}, \& {Zamorani}}]{smolcic08}
{Smol{\v c}i{\'c}}, V., {Schinnerer}, E., {Scodeggio}, M., {et~al.} 2008,
  \apjs, 177, 14

\bibitem[{{Smol{\v c}i{\'c}} {et~al.}(2010){Smol{\v c}i{\'c}}, {Zamorani},
  {Schinnerer}, {Bardelli}, {Bondi}, {B{\^i}rzan}, {Carilli}, {Ciliegi},
  {Elvis}, {Impey}, {Koekemoer}, {Merloni}, {Paglione}, {Salvato}, {Scodeggio},
  {Scoville}, \& {Trump}}]{smolcic10}
{Smol{\v c}i{\'c}}, V., {Zamorani}, G., {Schinnerer}, E., {et~al.} 2010, \apj,
  708, 909

\bibitem[{{Strateva} {et~al.}(2001){Strateva}, {Ivezi{\'c}}, {Knapp},
  {Narayanan}, {Strauss}, {Gunn}, {Lupton}, {Schlegel}, {Bahcall}, {Brinkmann},
  {Brunner}, {Budav{\'a}ri}, {Csabai}, {Castander}, {Doi}, {Fukugita}, {Gy{\H
  o}ry}, {Hamabe}, {Hennessy}, {Ichikawa}, {Kunszt}, {Lamb}, {McKay},
  {Okamura}, {Racusin}, {Sekiguchi}, {Schneider}, {Shimasaku}, \&
  {York}}]{strateva01}
{Strateva}, I., {Ivezi{\'c}}, {\v Z}., {Knapp}, G.~R., {et~al.} 2001, \aj, 122,
  1861

\bibitem[{{Vajgel} {et~al.}(2014){Vajgel}, {Jones}, {Lopes}, {Forman},
  {Murray}, {Goulding}, \& {Andrade-Santos}}]{vajgel14}
{Vajgel}, B., {Jones}, C., {Lopes}, P.~A.~A., {et~al.} 2014, \apj, 794, 88

\bibitem[{{van de Weygaert} \& {Icke}(1989)}]{icke89}
{van de Weygaert}, R. \& {Icke}, V. 1989, \aap, 213, 1

\bibitem[{{Velander}(2012)}]{velander12}
{Velander}, M.~B.~M. 2012, PhD thesis, Leiden Observatory, Leiden University,
  PO Box 9513, 2300 RA, Leiden, the Netherlands

\bibitem[{{Venturi} {et~al.}(1997){Venturi}, {Bardelli}, {Morganti}, \&
  {Hunstead}}]{venturi97}
{Venturi}, T., {Bardelli}, S., {Morganti}, R., \& {Hunstead}, R.~W. 1997,
  \mnras, 285, 898

\bibitem[{{Venturi} {et~al.}(2000){Venturi}, {Bardelli}, {Morganti}, \&
  {Hunstead}}]{venturi00}
{Venturi}, T., {Bardelli}, S., {Morganti}, R., \& {Hunstead}, R.~W. 2000,
  \mnras, 314, 594

\bibitem[{{Venturi} {et~al.}(2002){Venturi}, {Bardelli}, {Zagaria}, {Prandoni},
  \& {Morganti}}]{venturi02}
{Venturi}, T., {Bardelli}, S., {Zagaria}, M., {Prandoni}, I., \& {Morganti}, R.
  2002, \aap, 385, 39

\bibitem[{{Venturi} {et~al.}(2001){Venturi}, {Bardelli}, {Zambelli},
  {Morganti}, \& {Hunstead}}]{venturi01}
{Venturi}, T., {Bardelli}, S., {Zambelli}, G., {Morganti}, R., \& {Hunstead},
  R.~W. 2001, \mnras, 324, 1131

\end{thebibliography}

	\appendix
	
\end{document}